\documentclass[reqno,11pt]{amsart}
\usepackage{amsmath,amssymb,graphicx,a4wide,enumerate,url}
\usepackage[small,bf]{caption} 
\usepackage{subcaption}

\begin{document}
\parskip.9ex

\title[Comparison of Langevin Integrators for Coarse-Grained Polymer Simulations]
{Comparison of Modern Langevin Integrators for Simulations of Coarse-Grained Polymer Melts}

\author[J. Finkelstein]{Joshua Finkelstein}
\address[Joshua Finkelstein]
{Department of Mathematics \\ Temple University \\
1805 North Broad Street \\ Philadelphia, PA 19122}
\email{joshua.finkelstein@temple.edu}

\author[G. Fiorin]{Giacomo Fiorin}
\address[Giacomo Fiorin]
{Institute for Computational Molecular Science \\ Temple University \\
1805 North Broad Street \\ Philadelphia, PA 19122
\hspace{2em}and\hspace{2em}
National Heart, Lung and Blood Institute \\ Bethesda, MD}
\email{giacomo.fiorin@temple.edu}

\author[B. Seibold]{Benjamin Seibold}
\address[Benjamin Seibold]
{Department of Mathematics \\ Temple University \\
1805 North Broad Street \\ Philadelphia, PA 19122}
\email{seibold@temple.edu}
\urladdr{http://www.math.temple.edu/\~{}seibold}


\begin{abstract}
For a wide range of phenomena, current computational ability does not always allow for fully atomistic simulations of high-dimensional molecular systems to reach time scales of interest. Coarse-graining (CG) is an established approach to alleviate the impact of computational limits while retaining the same algorithms used  in atomistic simulations.
It is of importance to understand how algorithms such as Langevin integrators perform on non-trivial CG molecular systems, and in particular how large of an integration time step can be used without introducing unacceptable amounts of error into averaged quantities of interest. To investigate this, we examined three different Langevin integrators on a CG polymer melt: the recently developed BAOAB method by Leimkuhler and Matthews \cite{lk_article}, the Gr{\o}nbech-Jensen and Farago method \cite{gjf}, or G--JF, and the frequently used Br\"{u}nger-Brooks-Karplus integrator \cite{bbk}, also known as BBK. We compute and analyze key statistical properties for each. Our results indicate that the three integrators perform similarly when using a small friction parameter; however, outside of this regime the use of large integration steps produces significant deviations from the predicted diffusivity and steady-state distributions for all integration methods examined with the exception of G--JF. 
\end{abstract}

\maketitle

\section{Introduction}
A central obstacle in using molecular dynamics (MD) simulations for quantitative predictions in material science and molecular biology is the presence of a wide range of time scales that are not well-separated.
To investigate phenomena in such systems that occur over large time scales, while accurately resolving smaller ones, a large number of integration steps is required. For example, in fully atomistic simulations, the size of the integration time step is constrained by the fastest physical time scales and is typically on the order of femtoseconds. It becomes computationally inefficient, yet necessary, to use these relatively small time steps for integrating the medium to long-range time scale portion of the force field.

There exist several methods for working around this bottleneck. One such method is Nose-Hoover chains (with or without RESPA) \cite{tuckerman,tuckerman_respa,NH_chains}. This allows one to match the time step size for each degree of freedom to its corresponding time scale and compose the resulting operators via a symmetric Trotter splitting. Unfortunately, the Nose-Hoover chain equations result in the introduction of many undetermined parameters and, depending on the complexity of the system being simulated, may require different atoms to be coupled to different thermostats, thus complicating implementation. Other deterministic approaches which attempt to constrain the fast degrees of freedom are also used \cite{SHAKE}.

A Langevin thermostat is a simple and efficient way of simulating constant temperature conditions and has the intuitive physical interpretation of describing a molecular system in the presence of an implicit solvent or heat bath. The interaction between the heat bath and the system is collapsed into the friction parameter $\gamma$, thus avoiding the need to represent the heat bath as a set of particles altogether. In the context of coarse-grained (CG) models with explicit solvent, the missing microscopic degrees of freedom can be described by such stochastic forces, at least to a first approximation. For example, the dissipative-particle-dynamics (DPD) methodology \cite{dpd_chapter} relies on this description to systematically build coarse-grained models with low computational cost.

Because improvements in computer hardware are being introduced more slowly in recent years, the usefulness of MD models that run at speeds higher than atomistic simulations will also increase. As CG models become more commonplace, there is a need to systematically understand the numerical performance and the limitations of current temporal integration schemes on CG systems, avoiding the reliance on rules of thumb that were derived primarily for atomistic simulations.

In this work we examine methods used to numerically solve the Langevin equation, such as the ones by Leimkuhler and Matthews \cite{lk_article} and Gr{\o}nbech-Jensen and Farago \cite{gjf}, known as G--JF, with particular concern toward their respective sampling properties in CG simulations. Among the family of integrators described by Leimkuhler and Matthews \cite{lk_article}, the BAOAB method is the one characterised by the smallest configurational sampling error, and the only one here considered. Both the G-JF and BAOAB methods are weakly second-order accurate \cite{gjf_num_analysis,lk_article} and produce the exact configurational mean, variance and co-variance of the harmonic oscillator. This important property is not produced by many other Langevin schemes \cite{wang_skeel,bbk88}. To compare the two schemes with a representative of more traditional integration methods, our analysis includes the well-established Br\"{u}nger-Brooks-Karplus method \cite{bbk}, or BBK. Out of the many formulations of BBK \cite{lk} we chose one (indicated here as BBK$^*$) that performs well for the system under study: comparisons to the classical formulation are also made.

Several numerical studies have been conducted on the performance of these integrators in atomistic simulations (e.g.\ see \cite{gjf_test,baoab_test}). However, CG models tend to use smoother potential energy functions than their fully atomistic counterparts, and thus allow for much larger integration time steps. A key question that arises then is: how large can the time step $h$ be made without introducing unacceptable levels of error into averaged static and dynamic quantities? In \cite{gjf_CG_test}, G--JF is used to simulate a CG lipid bilayer in implicit solvent and averaged energy terms (both potential and kinetic) are examined: however, the system was simulated for relatively short MD trajectories ($<$50,000 steps) and distributions were not examined. Also included in that study was the Schneider-Stoll Langevin integrator \cite{SS}, the default option in LAMMPS \cite{LAMMPS} and ESPResSo \cite{espresso}. We elected not to include this integrator in our work, because small $\gamma h$ values are explicitly required in its derivation. Our present work considers a wide range of values for $\gamma h$. To date, we are not aware of any CG studies for BAOAB.

There are several choices of CG-model resolutions to choose from: a recent survey of several CG models \cite{grest} suggested that an adequate representation of the phase behavior seen in atomistic simulations is given by models of polyethylene chains with three or four methylene groups per CG particle. The model by Klein and coworkers \cite{shelley, sdk}, the MARTINI \cite{martini} and Salerno-Grest \cite{grest} models are significant examples of this level of resolution. Many other such models also exist \cite{cooke2005tunable,goetz1998computer,izvekov2005multiscale,brannigan2006implicit,orsi2008quantitative,murtola2009systematic,hadley2010structurally}. It is also recognised that technological requirements will motivate further effort to develop accurate CG models at lower levels of resolution (mapping into fewer CG particles for the same system). Transitioning into such models affects significantly the balance between Hamiltonian, stochastic and inertial terms in the equations of motion, and may require the use of existing Langevin methods outside of their typical range of parameter values.

In fact, Langevin parameters near the high-friction limit are of particular concern. Here, the acceleration may be neglected and the second-order integrator can safely be substituted with a first-order one for improved numerical stability. However, due to large differences in how existing CG models are formulated, as well as differences between potential energy terms in the same model, it is far from unusual to see applications of a Langevin thermostat that approach this region at least transiently. Unfortunately, any resulting biases in sampling are often difficult to detect due to the heterogeneous nature of the physical system examined, or its proximity to a phase transition.

Here we used as benchmark system a polyethylene melt ($\mathrm{C}_{48}$) modeled with three methylene groups per CG particle. All three schemes, BAOAB, BBK and G--JF, were compared for a wide-range of friction parameter values and time step sizes by examining relevant statistical quantities from the simulations. The key finding of our study is that in the high-friction ($\gamma \approx .1 \text{ fs}^{-1}$) regime, the G--JF method performs measurably better than BAOAB and BBK in reproducing molecular diffusivity and configurational distributions.
The results obtained provide indications that Langevin integrators with similar properties to G--JF should be considered for use in CG simulations that aim to preserve dynamic properties and stationary distributions equally accurately. Though, diffusivity notwithstanding, BAOAB and G--JF sample equally well the configurational distributions of the system considered here.

\section{Background and Theory}
We consider an $N$-particle system with potential energy $U$, immersed in a heat bath with the constant temperature $T$, modeled by the Langevin equation:
\begin{align}
\begin{split}
    d\mathbf{Q} &=  \mathbf{M}^{-1} \mathbf{P} dt\;, \\
    d\mathbf{P} &= -\mathbf{M}^{-1} \nabla U(\mathbf{Q}) dt  -\gamma \mathbf{M}^{-1} \mathbf{P}dt + \sigma \mathbf{M}^{-1/2} d\mathbf{W}\;.
\end{split}
\label{eq:langevin}
\end{align}
Here $\sigma = \sqrt{2 k_b T \gamma}$ is the noise coefficient, $k_b$ is Boltzmann's constant, $\gamma$ is the (spatially independent) collision rate parameter (measured in units of fs$^{-1}$), $\mathbf{M}$ a diagonal mass matrix, $\mathbf{W}$ is $3N$-dimensional Brownian motion, and $\mathcal{H}(\mathbf{Q},\mathbf{P})= \mathbf{P}^T \mathbf{M}^{-1} \mathbf{P} + U(\mathbf{Q})$ is the Hamiltonian. The Langevin equation is a stochastic differential equation (SDE), so we use capital letters to remind us that position and velocity are stochastic processes.

Usually in MD, it is not the exact dynamics generated by (\ref{eq:langevin}) that are of interest, but rather an accurate sampling of the equilibrium distributions in phase space. Assuming that $\mathcal{H}(q,p)$ is such that $e^{-\mathcal{H}(q,p)}$ is integrable, one expects (\ref{eq:langevin}) to be ergodic and have the stationary distribution
\[
\mu(dq dp) = \mathcal{Q}^{-1} e^{-\mathcal{H}(q,p)/k_b T} \ dqdp\;,
\]
where $\mathcal{Q}$ is a normalisation constant and $\mu$ is the Boltzmann-Gibbs, or canonical, distribution. For realistic MD potentials, such as Lennard-Jones and/or Coulombic interaction forces, the solution to (\ref{eq:langevin}) needs to be approximated numerically. Moreover, in many applications, such as the one here, $U$ is non-globally Lipschitz and singular. Consequently, many standard results in SDE theory do not apply, thus limiting the possibilities of a complete formal analysis of numerical schemes for (\ref{eq:langevin}). One must therefore study these schemes computationally.

The fidelity to which numerical schemes reproduce the Boltzmann-Gibbs distribution $\mu$ is our principal interest. It is the hope that infinite time-averaged observables obtained from these numerical schemes would reproduce correct statistical averages with respect to $\mu$. However, even with ergodicity typically assumed, the steady states produced by each numerical scheme, $\mu_{\text{G--JF}}$, $\mu_{\text{BAOAB}}$, and $\mu_{\text{BBK}}$, respectively, will in general differ from $\mu$, and depend on the friction parameter $\gamma$ and time step $h$. Therefore we expect two distinct sources of error in the calculation of distributions and observables: use of a finite trajectory instead of an infinite one and the error due to the numerical scheme's steady state distribution differing from the canonical distribution.

Although, in general, statistical averages generated by numerical approximations of (\ref{eq:langevin}) cannot be derived analytically, exact formulas for mean, variance and correlation can be calculated in the flat and harmonic potential case, which is enough to characterise any stationary distribution when starting with Gaussian initial conditions.

\subsection{Numerical Methods Studied}
In this work, we examined three different Langevin integration schemes: G--JF, BAOAB, and the Br\"{u}nger-Brooks-Karplus method \cite{bbk}, known as BBK, on a CG polymer melt. Both G--JF and BAOAB are included as options in LAMMPS and NAMD, respectively \cite{lammps_source_code,namd_source_code}. The G--JF thermostat is a stochastic two-stage partitioned Runge-Kutta method \cite{reverse_leapfrog,gjf_num_analysis} and was shown to have highly desirable configurational properties \cite{gjf}; particularly, Einstein's diffusion relation holds exactly and the configurational averages for the harmonic oscillator are independent of both the time step $h$ and the friction parameter $\gamma$.

As described in \cite{lk_article}, BAOAB is but one of many splitting schemes obtained by composing solution operators in various orderings which evolve the A, B and O portions of the Langevin vector field (\ref{eq:langevin}):
\begin{align*}
    \begin{pmatrix} d\mathbf{Q}\\d\mathbf{P} \end{pmatrix} &=  \underset{A}{\underbrace{\begin{pmatrix} \mathbf{M}^{-1} \mathbf{P} \\0  \end{pmatrix} }}dt +  \underset{B}{\underbrace{\begin{pmatrix} 0 \\ -\mathbf{M}^{-1} \nabla U(\mathbf{Q})   \end{pmatrix} }}dt + \underset{O}{\underbrace{\begin{pmatrix}  0 \\-\gamma \mathbf{M}^{-1} \mathbf{P} dt + \sigma \mathbf{M}^{-1/2}  d\mathbf{W}  \end{pmatrix}}}  \;.
\end{align*}
BAOAB is the result of taking half steps for B and then A, a full step for O and then half steps again for A and then B.

Similar to G--JF, BAOAB also reproduces exact sampling for the harmonic oscillator. Moreover, it possesses an additional favorable configurational sampling property, termed ``super-convergence'' \cite{lk_article}: when $\gamma$ is sufficiently large, the leading order error terms of averaged phase space quantities will exhibit a $4$th order error scaling in $h$ for typical time step values, thereby yielding more accurate averages with the same computational effort. This property provides some motivation as to why BAOAB is of interest to practitioners.

BBK has been a well-known Langevin discretisation method for the last three decades and is the default Langevin integrator in the popular MD suite, NAMD \cite{namd_source_code}. Similar to BAOAB, BBK is also a splitting method. It is weakly first order accurate \cite{bbk88} and in the free particle case reproduces the Einstein relation. However, exact statistics are not recovered in the case of the harmonic oscillator, in sharp contrast to G--JF and BAOAB.
Equation (\ref{eq:langevin}) is often re-formulated in terms of position and velocity, instead of momentum:
\begin{align}
\begin{split}
    d\mathbf{Q} &=  \mathbf{V} dt\;, \\
    d\mathbf{V} &= -\mathbf{M}^{-1} \nabla U(\mathbf{Q}) dt  -\gamma \mathbf{V}dt + \sigma \mathbf{M}^{-1/2} d\mathbf{W}\;.
\end{split} \label{eq:langevin_velocity}
\end{align}
This form will serve as the governing equation for our forthcoming analysis and discussion. Set $a:=(1-\gamma h/2)(1+\gamma h/2)^{-1}$ and $b:=(1+\gamma h/2)^{-1}$, with $h$ being the time step used for discretisation. Note that, for $\gamma h$ sufficiently small, one has
\begin{align*}
a:=\frac{1-\gamma h/2}{1+\gamma h /2}
=e^{-\gamma h} +O((\gamma h)^3)\;,
\end{align*}
which leads to $a^n = e^{-\gamma t} + O((\gamma h)^2)$ when $n h = t$. The quantities $a$ and $a^n$ appear several times in the subsequent paragraphs and sections.

We start by displaying the recursion formulas for the considered numerical schemes discretising (\ref{eq:langevin_velocity}). For a fixed time step $h>0$ and initial configuration of $(Q_0,V_0)$, the position and velocity at time $t=nh$ for each scheme are displayed below. Define $\widetilde{\sigma} = \sqrt{k_b T (1-e^{-2 \gamma h})}$. Then,
the G--JF update rule is:
\begin{align}
\begin{split}
\textbf{Q}_{n} & = \textbf{Q}_{n-1} + b h \textbf{V}_{n-1} - \frac{b h^2}{2} \textbf{M}^{-1}\nabla U(\textbf{Q}_{n-1}) + \frac{b \sigma h^{3/2}}{2} \textbf{M}^{-1/2} \boldsymbol{\xi}_{n-1}\;, \\
\textbf{V}_{n} & = a \textbf{V}_{n-1} - \frac{h}{2} \textbf{M}^{-1} (a \nabla U(\textbf{Q}_{n-1}) + \nabla U(\textbf{Q}_{n})) + {b \sigma} \sqrt{h} \textbf{M}^{-1/2} \boldsymbol{\xi}_{n-1}\;,
\end{split}\label{eq:gjf}
\end{align}
the BAOAB method is:
\begin{align}
\begin{split}
\textbf{Q}_n &= \textbf{Q}_{n-1}+\frac{h}{2}(1+e^{-\gamma h}) \textbf{V}_{n-1}- \frac{h^2}{4}(1+e^{-\gamma h}) \textbf{M}^{-1} \nabla U(\textbf{Q}_{n-1}) + \frac{\tilde{\sigma} h}{2} \textbf{M}^{-1/2} \boldsymbol\xi_{n-1}\;, \\
\textbf{V}_n &= e^{-\gamma h} \textbf{V}_{n-1} - \frac{h}{2} \textbf{M}^{-1} (e^{-\gamma h} \nabla U(\textbf{Q}_{n-1})+\nabla U(\textbf{Q}_{n})) + {\tilde{\sigma}}  \textbf{M}^{-1/2} \boldsymbol\xi_{n-1}\;,
\end{split}\label{eq:baoab}
\end{align}
and finally the BBK method is:
\begin{align}
\begin{split}
\textbf{Q}_{n} &= \textbf{Q}_{n-1} + h (1-\gamma h /2) \textbf{V}_{n-1} - \frac{h^2}{2} \textbf{M}^{-1}\nabla U( \textbf{Q}_{n-1}) + \frac{\sigma {h}^{3/2}}{2} \textbf{M}^{-1/2} \boldsymbol \xi_{n-1}\;,\\
\textbf{V}_{n} & = a\textbf{V}_{n-1} - \frac{b h}{2}\textbf{M}^{-1}(\nabla U( \textbf{Q}_{n-1}) + \nabla U( \textbf{Q}_{n})) + \frac{b \sigma \sqrt{h}}{2} \textbf{M}^{-1/2}(\boldsymbol \xi_{n-1} + \boldsymbol \xi_{n})\;.
\end{split}\label{eq:bbk}
\end{align}
In its original formulation \cite{bbk}, the BBK numerical scheme is given for only the position, leaving some ambiguity as to how the velocities are defined. Using the second order approximation $\textbf{V}_{n} \approx (\textbf{Q}_{n+1}-\textbf{Q}_{n-1})/2h$, the splitting formulation of BBK for both position and velocity is obtained (e.g.\ as seen in \cite{langevin_stabilization}). This is the same substitution one employs in transforming the position-only Verlet integrator to velocity Verlet:
\begin{align}
\begin{split} \label{eq:bbk_splitting}
\textbf{V}_{n-1/2} &= \textbf{V}_{n-1} + \frac{h}{2}\textbf{M}^{-1} \big(-\nabla U(\textbf{Q}_{n-1}) - \gamma \textbf{M} \textbf{V}_{n-1} + \tfrac{\sigma}{\sqrt{h}} \textbf{M}^{1/2}  \boldsymbol \xi_{n-1} \big) \;, \\
\textbf{Q}_{n} &= \textbf{Q}_{n-1} + h \textbf{V}_{n-1/2} \;,\\
\textbf{V}_{n} &= \textbf{V}_{n-1/2} + \frac{h}{2}\textbf{M}^{-1} \big(-\nabla U( \textbf{Q}_{n}) - \gamma \textbf{M} \textbf{V}_{n} + \tfrac{\sigma}{\sqrt{h}} \textbf{M}^{1/2} \boldsymbol \xi_{n} \big) \;.
\end{split}
\end{align}
It is then a simple exercise to derive (\ref{eq:bbk}) from (\ref{eq:bbk_splitting}). In particular, this method requires two independent random variables $\boldsymbol \xi_{n-1}$ and $\boldsymbol \xi_{n}$ where $\boldsymbol \xi_{n}$ is then re-used in the next step. However, \cite{lk} suggests the use of several variations of BBK which vary in how these random variables are selected. One such variation, which we denote by BBK$^*$, is obtained by taking $\boldsymbol \xi_{n-1} = \boldsymbol \xi_{n}$ in (\ref{eq:bbk_splitting}), and not conducting any re-use in the next step. This BBK$^*$ variant is not equivalent to the original version of BBK; and in fact, numerical tests indicated better all-around performance with our particular CG molecular system of interest in the commonly used regime of $\gamma h \le .01$ (see Fig.~\ref{fig:bbk_vs_bbk*}).

\subsection{Analytical Properties of the Methods in One Dimension}
Before moving to the computational results, we first summarise the key statistical properties for each of the three schemes in one dimension. This section attempts to highlight some key structural differences (and similarities) of the three schemes. Some schemes reproduce certain statistical quantities exactly whereas others reproduce such quantities only approximately.

We consider the standard examples of a single particle diffusing in a heat bath, and a standard harmonic oscillator, modeling for instance a covalent bond between two particles. In these cases, we can directly compare statistical quantities generated by the numerical schemes with those generated by the true analytical solution of (\ref{eq:langevin_velocity}). Though simplistic, these examples illustrate some important properties. Most of the calculations for these examples have, in parts, been previously exposited \cite{reverse_leapfrog, gjf,bbk88,lk_article,wang_skeel,lk_article_2}.

\subsubsection{Harmonic Potential}
We consider a potential function of the form $U(Q) = \omega Q^2/2$, with $\omega>0$, so that $\mathcal{H}(Q,P) = P^2/2m+\omega Q^2/2$, and (\ref{eq:langevin_velocity}) becomes
\begin{align}
\begin{split}
    d{Q}  &=  {V} dt\;, \\
    d{V}  &=-\omega Q dt - \gamma V dt + \sigma {m}^{-1/2} dW\;.
    \label{eq:linear_langevin}
\end{split}
\end{align}
The following calculations show that the variance of position for the BAOAB and G--JF integrators is independent of $\gamma$ and $h$. This is not true for BBK. For the linear system (\ref{eq:linear_langevin}), stationary distributions can be analytically derived for the three methods, denoted by $\mu_{\text{BAOAB}}$, $\mu_{\text{BBK}^*}$, and $\mu_{\text{G--JF}}$, respectively. We re-write (\ref{eq:gjf}) applied to (\ref{eq:linear_langevin}) into matrix form:
\begin{align*}
	\begin{bmatrix}
	Q_{n+1}\\
	V_{n+1}
	\end{bmatrix}
	&=
    \begin{bmatrix}
	 1 - \frac{b \omega h^2}{2m} &  {bh}  \\
	   \frac{-h \omega b}{m}(1-\frac{h^2 \omega}{4m}) & a - \frac{h^2 \omega b}{2m}
	\end{bmatrix}
    \begin{bmatrix}
	Q_{n}\\
	V_{n}
	\end{bmatrix}
	+
	\begin{bmatrix}
	\frac{ b \sigma h^{3/2}}{2 \sqrt{m}} \\
	\frac{b \sigma \sqrt{h}}{\sqrt{m}} (1-\frac{h^2 \omega}{4m} )
	\end{bmatrix}  \xi_{n}\;. \label{eq:linear_gjf}
\end{align*}
This is a two-dimensional ergodic Markov chain with a unique stationary measure, $\mu_{\text{G--JF}}$. We can then calculate a corresponding matrix equation for $Q_{n+1}^2$, $V_{n+1}^2$ and $Q_{n+1}V_{n+1}$. Taking expectations on both sides of this equation, and taking $n\to \infty$, yields a subsequent $3 \times 3$ linear system for $\mathbb{E}(Q_\infty^2)$, $\mathbb{E}(V_\infty^2)$ and $ \mathbb{E}((QV)_\infty)$, the vector of steady-state averages. Solving the resultant linear system yields the G--JF stationary distribution:
\begin{align*}
\mu_{\text{G--JF}}(dq dv) &\propto \exp \bigg(-\beta\bigg(\frac{mv^2}{2(1-\tfrac{h^2 \omega}{4m})}  +\frac{\omega q^2}{2} \bigg) \bigg)dq dv \;.
\end{align*}
Using (\ref{eq:baoab}) and (\ref{eq:bbk}) one can derive analogous linear equations and expressions for BBK$^*$ and BAOAB, as follows.
\begin{itemize}
\item\textbf{BBK$^*$:}
\begin{align}
\begin{split}
	\begin{bmatrix}
	Q_{n}\\
	V_{n}
	\end{bmatrix}
	&=
    \begin{bmatrix}
	  1-\frac{h^2 \omega}{2m} & h(1-\frac{\gamma h}{2})  \\
	  \frac{-h b \omega}{m} \big( 1 - \frac{h^2 \omega}{4 m} \big) & a(1-\frac{h^2 \omega}{2m})
	\end{bmatrix}
    \begin{bmatrix}
	Q_{n-1}\\
	V_{n-1}
	\end{bmatrix}
	+
	\begin{bmatrix}
	\frac{\sigma h^{3/2}}{2\sqrt{m}} \xi_{n-1}\\
	\frac{\sigma b \sqrt{h}}{\sqrt{m}}(1-\frac{h^2\omega}{4m}) \xi_{n-1}
	\end{bmatrix} \;
\end{split}
\end{align}
and
\begin{align*}
\mu_{\text{BBK$^*$}}(dq dv) &\propto \exp \bigg(-\beta\bigg(\frac{mv^2}{2} + \frac{ (1-\tfrac{ h^2 \omega}{4m} )}{(1-\gamma h/2)^2} \frac{ \omega q^2}{2}  \bigg)  \bigg)dq dv\;.
\end{align*}
\item\textbf{BAOAB:}
\begin{align}
\begin{split}
	\begin{bmatrix}
	Q_{n+1}\\
	V_{n+1}
	\end{bmatrix}
	&=
    \begin{bmatrix}
	 1-(1+e^{-\gamma h}) \frac{h^2 \omega}{4m} & (1+e^{-\gamma h})\tfrac{h}{2}  \\
	  - (1-\tfrac{h^2 \omega}{4m})(1+e^{-\gamma h})\tfrac{h \omega}{2m} & e^{-\gamma h} (1-\tfrac{h^2 \omega}{4m})
	\end{bmatrix}
    \begin{bmatrix}
	Q_{n}\\
	V_{n}
	\end{bmatrix}\\
	& \hspace{2in}+
	\begin{bmatrix}
	 \frac{h}{2} \sqrt{\frac{k_b T}{m} (1-e^{-2 \gamma h})} \\
    (1-\tfrac{h^2 \omega}{4m})  \sqrt{\frac{k_b T}{m} (1-e^{-2 \gamma h})}
	\end{bmatrix}  \xi_{n}\;,
\end{split} \label{eq:linear_baoab}
\end{align}
and
\begin{align*}
\mu_{\text{BAOAB}}(dq dv) &\propto \exp \bigg(-\beta\bigg(\frac{mv^2}{2(1-h^2 \omega/4m)}  +\frac{ \omega q^2}{2}  \bigg)  \bigg)dq dv\;.
\end{align*}
\end{itemize}
The variances and covariances for position and velocity are listed in Table~\ref{table:harmonic_numerical_stats}. These expressions show the dependence on the dimensionless quantity $\gamma h$ and the time step $h$. In particular, both BAOAB and G--JF produce the exactly correct configurational variance and co-variance.

\begin{table}[ht]
\centering
 \begin{tabular}{|c||c|c|c|}
 \hline
 Method & $\langle Q_n^2 \rangle_{h,\gamma} $ &  $\langle V_n^2 \rangle_{h,\gamma}$ &  $\langle Q_nV_n \rangle_{h,\gamma}$ \\[1ex]
 \hline\hline
 Exact & $\frac{k_b T}{ \omega}$ & $\frac{k_bT}{m}$ & 0  \\[1ex]
 \hline
 G--JF & $\frac{k_bT}{\omega}$ &  $\frac{k_bT}{m}(1-h^2 \omega /4m)$ &  0  \\[1ex]
 BBK$^*$ & $\frac{k_b T}{ \omega} { (1-\frac{ h^2 \omega}{4m} )} (1-\gamma h/2)^{-2} $ & $\frac{k_bT}{m}$ & 0  \\ [1ex]
 BAOAB & $\frac{k_bT}{\omega}$ & $\frac{k_bT}{m} (1-h^2 \omega/4m)$ & 0  \\[1ex]
 \hline
 \end{tabular}
\vspace{.3em}
\caption{Numerical stationary averages, as functions of $h$ and $\gamma$, for the three numerical methods applied to the one-dimensional harmonic oscillator. Both G--JF and BAOAB are exact for position variance, while BBK$^*$ is not.}
\label{table:harmonic_numerical_stats}
\end{table}

\subsubsection{Thermal Diffusion}
The simplest possible case for (\ref{eq:langevin_velocity}) is when $F \equiv 0$, i.e., free diffusion. Albeit simple, it is insightful to understand the behavior of the integrators in this case. In thermal diffusion, (\ref{eq:langevin_velocity}) reduces to
\begin{align*}
\begin{split}
dQ &= V dt\;,\\
dV &= -\gamma V dt + \sigma m^{-1/2} dW\;,
\end{split}
\end{align*}
which can be solved analytically. The velocity $V_t$ is an Ornstein-Uhlenbeck process and has solution:
\begin{align}
V_t = e^{-\gamma t} V_0 + \frac{\sigma}{\sqrt{m}} \int_0^t e^{-\gamma(t-s)} dW_s\;. \label{eq:brownian_v_true}
\end{align}
The particle position, $Q_t$, is then
\begin{align}
\begin{split}
Q_t &= Q_0 + \frac{1}{m} \int_0^t V_u du = Q_0 + \tfrac{1}{\gamma}(1-e^{-\gamma t}) V_0 + \frac{\sigma}{\sqrt{m}} \int_0^t \bigg( \int_0^u e^{-\gamma(u-s)} dW_s \bigg) du\;.
\end{split} \label{eq:brownian_q_true}
\end{align}
Equation (\ref{eq:brownian_q_true}) is used to find the mean position and mean squared position. Assuming $\mathbb{E} (Q_t)=0$,
\begin{align*}
\mathbb{E}(Q_t^2) &= \mathbb{E}(Q_0^2) +  2\overbrace{\mathbb{E}(Q_0)}^{=0} \mathbb{E}\bigg(\int_0^t V_u du \bigg) +\mathbb{E} \bigg( \int_0^t V_u du \bigg)^2\\
&= \mathbb{E}(Q_0^2) + \frac{1}{\gamma^2}(1-e^{-\gamma t})^2 \mathbb{E}(V_0^2) + 2D \bigg(t-\frac{2}{\gamma}(1-e^{-\gamma t}) + \frac{1}{2 \gamma}(1-e^{-2\gamma t})  \bigg)\;,
\end{align*}
where $D:=k_b T/m\gamma$ is the diffusion coefficient. The large time asymptotic behavior of the mean squared position for mean zero initial position is then
\begin{align}
\mathbb{V}(Q_t) \sim 2Dt \;.
\label{eq:brownian_true_variance}
\end{align}
The velocity auto-correlation function and the covariance can also be computed from (\ref{eq:brownian_v_true}) and (\ref{eq:brownian_q_true}). These quantities are displayed in Table~\ref{table:numerical_and_analytic_stats}.

\subsubsection{Diffusive Behavior of the Numerical Schemes}
We set $\omega = 0$ in (\ref{eq:linear_gjf})--(\ref{eq:linear_baoab}) to obtain the update rules for each numerical scheme in the zero potential case. A key quantity of interest is the mean square displacement for the particle position. For BBK$^*$,
\begin{align}
\begin{split}
Q_{n} &= Q_{n-1}
+ h(1-\gamma h/2) V_{n-1} + \frac{\sigma h}{2 \sqrt{m}} \xi_{n-1}\;, \\
V_{n} &= a V_{n-1} + \frac{b \sigma \sqrt{h}}{ \sqrt{m}} \xi_{n-1} \;.
\end{split}\label{eq:brownian_bbk}
\end{align}
Given a fixed time step size $h>0$, iterate the above recursive formula backwards to write the position as a finite sum of independent Gaussians and the initial conditions:
\begin{align*}
Q_n = Q_0 + \frac{(1-\tfrac{\gamma h}{2})h(1-a^{n+1})}{1-a} V_0 + \frac{\sigma  h}{\sqrt{m}} \sum_{k=0}^{n-1} \xi_k \bigg( \frac{a(1-a^{k+1})}{1-a} + \frac{1}{2} \bigg)\;.
\end{align*}
Therefore,
\begin{align*}
\begin{split}
\mathbb{V}(Q_n) =  \mathbb{V}(Q_0) + & (1-\gamma h/2)^2 h^2 \bigg(\frac{ 1-a^{(n+1)}}{1-a} \bigg)^2\mathbb{V}(V_0) \\
& + 2D \bigg( t - \frac{2}{\gamma}(1-a^n)  (1-\gamma h/2)^2 + \frac{1}{2\gamma}(1-a^{2n})  (1-\gamma h/2)^4 \bigg)\;.
\end{split}
\end{align*}
where $t=nh$. Sending $n \to \infty$ shows that the scheme preserves the Einstein diffusion relation in the limit. As with BBK$^*$, G--JF preserves the Einstein diffusion relation in the long time limit as well. The position and its mean square displacement at time $t=nh$ are:
\begin{align*}
Q_n &= Q_0 + \frac{bh(1-a^{n+1})}{1-a} V_0 + \frac{\sigma b h^{3/2}}{\sqrt{m}} \sum_{k=0}^{n-1} \xi_k \bigg( \frac{(1-a^{k+1})b}{1-a} + \frac{1}{2} \bigg)\;,\\
\mathbb{V}(Q_n) &= \mathbb{V}(Q_0) + \frac{b^2h^2(1-a^{(n+1)})^2}{(1-a)^2}\mathbb{V}(V_0) + 2D \bigg( t - \frac{2}{\gamma}(1-a^n) a + \frac{1}{2\gamma}(1-a^{2n}) a^2 \bigg)\;,
\end{align*}
which has the same limiting behavior as in \cite{gjf}. Additionally, both BBK$^*$ and G--JF generate the correct steady-state behavior for velocity. For a fixed time step $h$, the BAOAB scheme approximates thermal diffusion via:
\begin{align*}
Q_{n} &=  Q_{n-1} + \tfrac{h}{2}(1+e^{-\gamma h}) V_{n-1} + \tfrac{h}{2}\sqrt{\tfrac{k_b T (1-e^{-2 \gamma h})}{m}} \xi_{n-1}\\
V_{n} &= e^{-\gamma h} V_{n-1} + \sqrt{\tfrac{k_b T (1-e^{-2 \gamma h})}{m}} \xi_{n-1} \;.
\end{align*}
An important distinction here is that $V_{n+1}$ is given by the exact Ornstein-Uhlenbeck flow (in law), whereas the velocity updates are only approximate for BBK$^*$ and G--JF. As done with BBK$^*$ and G--JF, by iterating backwards, we can write the position as a finite sum of independent identically distributed random variables:
\begin{align}
\begin{split}
Q_n = Q_0 + \frac{h}{2} &\bigg(\frac{1+e^{-\gamma h}}{1-e^{-\gamma h}} \bigg) (1-e^{-\gamma t}) V_0  \label{eq:BAOAB_var} \\
&+ \frac{h}{2}\frac{k_b T}{m}(1-e^{-2\gamma h}) \sum_{k=0}^{n-1} \xi_k \bigg( \frac{1+e^{-\gamma h}}{1-e^{-\gamma h}}(1-e^{-\gamma h(k+1)}) + 1 \bigg).
\end{split}
\end{align}
Then $\mathbb{V}(Q_n)$ is
\begin{align*}
&\mathbb{V}(Q_0) + \frac{\gamma h}{4}\bigg( \frac{1+e^{-\gamma h}}{1-e^{-\gamma h}} \bigg)^2 (1-e^{-\gamma t})^2 \mathbb{V}(V_0) \\
&+2D \bigg( \frac{\gamma h}{2} \frac{(1+e^{-\gamma h})^2}{1-e^{-2 \gamma h}} t - \frac{\gamma h^2}{2} \frac{( 1-e^{-2\gamma h})(1+e^{-\gamma h})}{(1-e^{-\gamma h})^3}(1-e^{- \gamma t}) e^{-\gamma h} \\
& \hspace{3in}+ \frac{1}{2\gamma}(1-e^{-2 \gamma t}) e^{-2 \gamma h} \bigg).
\end{align*}
This appears to be vastly different than the other schemes, but one can check that (\ref{eq:brownian_true_variance}) is recovered when sending $\gamma h \to 0$. So whenever $\gamma$ and $h$ are fixed such that $\gamma h$ is sufficiently small, $\lim_{n \to \infty} \mathbb{V}(Q_n)/nh \approx 2D$, and BAOAB produces an acceptable approximation to the correct diffusive behavior. More particularly, the time evolution of the variance for BAOAB at large times evolves according to $2\widetilde{D}t$ where
\[
\widetilde{D}=D \bigg( \frac{\gamma h}{2} \frac{(1+e^{-\gamma h})^2}{1-e^{-2 \gamma h}} \bigg) \;,
\]
is the effective BAOAB diffusion coefficient, showing how the calculated diffusion for BAOAB deviates from theory when $\gamma h$ is sufficiently large. The quantity inside the parentheses tends to 1 as $\gamma h\to 0$ but exhibits linear behavior as $\gamma h$ is increased. Key statistical quantities for each numerical scheme are tabulated in Table~\ref{table:numerical_and_analytic_stats} for simple initial conditions.
In fact, among all methods within the aforementioned A,B,O family of splitting schemes, when a single random sample per time step is desired, the incorrect diffusive behavior elucidated above is universal. In the free particle case, B induces the identity operator, so that the number of possible lettered combinations reduce to AO, OA, and AOA. A quick calculation reveals that none of these methods reproduces the Einstein relation.

\begin{table}
   \centering
   \begin{tabular}{|c||c|c|c|c|}
      \hline
      Method & $  \langle Q_n^2 \rangle_{h,\gamma} $ &  $\langle V_n^2
      \rangle_{h,\gamma}$ &  $\langle Q_nV_n \rangle_{h,
      \gamma}$ & $C_{h,\gamma}(nh)$ \\[1ex]
      \hline\hline
      Exact & $\sim 2 Dnh $ & $k_b T/m$ & $(1-e^{-\gamma h n})D $ & $ e^{-\gamma hn}\tfrac{k_bT}{m} $ \\[1ex]
      \hline
      G--JF & $\sim 2Dnh$ &  $k_bT/m$ &  $(1-a^n) D$ & $a^n \tfrac{k_b T}{m}$ \\
      BBK$^*$ & $\sim 2Dnh$ & $k_bT/m$ & $(1-a^n) D \big(1 + \tfrac{\gamma h}{2} \big)$ & $a^n \tfrac{k_b T}{m}$ \\[1ex]
      BAOAB & $ \sim ({\gamma h})  \big( \frac{1+e^{- \gamma h}}{1-e^{- \gamma h}} \big) Dnh $ & $k_b T/m$ &
      $(1-e^{-\gamma h n})D \big(\tfrac{\gamma h}{2} \tfrac{1+e^{-\gamma h}}{1-e^{-\gamma h}} \big)$ & $
      e^{-\gamma h n}\tfrac{k_bT}{m} $ \\[1ex]
      \hline
   \end{tabular}
\vspace{.3em}
\caption{The analytic and numerical schemes' averages in the case of Brownian motion (Einstein diffusion) with $\delta_0$--distributed $Q_0$ and Maxwell-Boltzmann-distributed $V_0$. The averages for the numerical schemes are given as functions of $h$ and $\gamma$. Both G--JF and BBK$^*$ are exact for position variance, while BAOAB is not. $C_{h,\gamma}(nh)$ is the velocity autocorrelation function of the numerical scheme with time step $h$ and friction parameter $\gamma$ at time $nh$.}
\label{table:numerical_and_analytic_stats}
\end{table}

\section{Computational Methodology}
In our computational study we considered a collection of 128 poly-ethylene chains ($\text{C}_{48}\text{H}_{98}$), simulated at a fixed temperature of 450~K (i.e., well above the melting point) in a box with side lengths of $58.065\text{~\AA}$ and periodic boundary conditions.
This resulted in a density of $0.4415\text{~amu} / \text{\AA}^3 = 0.7331$~g/ml. We used a coarse-grained model for the polymer melt obtained directly from liquid-phase physical properties \cite{sdk}. The hydrocarbon chains are modeled in coarse-grained resolution, where the $-$CH2CH2CH2$-$ and CH3CH2CH2$-$ groups are mapped to spherical ``beads", called CM and CT, respectively \cite{sdk}. Compared to atomistic resolution, this reduces the system from a total of 18,432 atoms to 2,048 CG beads. Initial positions of the CM and CT particles are taken from their respective centers-of-mass, and are evolved according to the following interaction energy:
\begin{align*}
U(\mathbf{q}) = \sum_{a \in \text{angles}} k_a (\theta - \theta_0)^2& + \sum_{b \in \text{bonds}} k_\text{b} (r - r_0)^2 +  U_{LJ}^{({\alpha,\beta})} (\mathbf{q})\;,
\end{align*}
where $({\alpha,\beta}) \in \{(\text{CT},\text{CT}), (\text{CT},\text{CM}), (\text{CM},\text{CM})\}$ and $U_{LJ}^{({\alpha,\beta})}$ is the 9--6 Lennard-Jones potential \cite{sdk}
\begin{align*}
U_{LJ}^{({\alpha,\beta})}(\mathbf{q}) = \sum_{i \neq j}^N \frac{27}{4} \epsilon_{\alpha,\beta}  \bigg( \bigg(\frac{\sigma_{\alpha,\beta}}{|\mathbf{q}_i-\mathbf{q}_j| }\bigg)^9 - \bigg( \frac{\sigma_{\alpha,\beta}}{|\mathbf{q}_i-\mathbf{q}_j| } \bigg)^6 \bigg) \cdot 1_{\{|\mathbf{q}_i-\mathbf{q}_j| < \delta\}}\;,
\end{align*}
with $\epsilon_{\text{CT,CT}} = 0.42\text{~kcal/mol}$ and $\sigma_{\text{CT,CT}} = 4.506\text{~\AA}$ for the CT$-$CT interaction,\linebreak $\epsilon_{\text{CT,CM}}= 0.444\text{~kcal/mol}$ and $\sigma_{\text{CT,CM}} = 4.5455\text{~\AA}$ for the CT$-$CM interaction, and $\epsilon_{\text{CM,CM}} = 0.469\text{~kcal/mol}$ and $\sigma_{\text{CM,CM}} = 4.585\text{~\AA}$ for the CM$-$CM interaction. Here $\delta$ represents the Lennard-Jones cut-off distance of 15~\AA. The function $1_{\{|\mathbf{q}_i-\mathbf{q}_j| < \delta\}}$ is defined to be 1 for all pairs $i,j$ satisfying $|\mathbf{q}_i-\mathbf{q}_j| < \delta$ and 0 otherwise. The CM$-$CM bonds have force constant $k_b=6.16$~kcal/mol and equilibrium length $r_0 = 3.64 \text{ \AA}$, and the CM$-$CT bonds have force constant $k_b=6.16$~kcal/mol and equilibrium length $r_0 = 3.65\text{~\AA}$. The force constants for the CM$-$CM$-$CM and CM$-$CM$-$CT angles were the same value: $k_a = 1.19$~kcal/mol$\cdot\text{rad}^2$, equilibrium angles were $\theta_0 = 173^\circ$ and $\theta_0 = 175^\circ$, respectively.

\begin{figure}
\centering
\includegraphics[width=.45\textwidth]{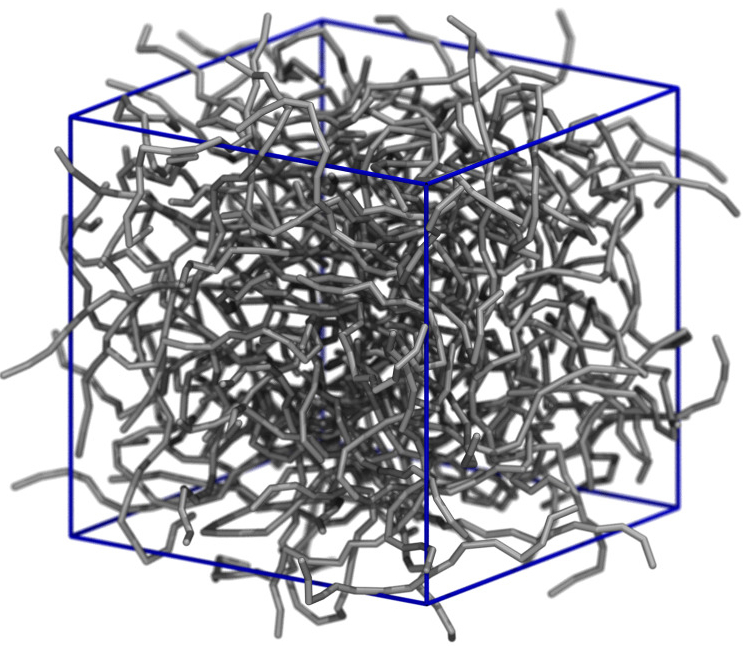} 
\caption{Graphical representation of the polymer melt inside the unit cell (blue): 128 CG $\text{C}_{48}$ polymer chains. This rendering was made using VMD after an initial thermalisation run.}
\end{figure}
To construct a reference ensemble, we first identified a value for the integration time step that was guaranteed not to introduce artifacts. The fastest CG bond oscillation is of the type CM$-$CM. The mass of a CM particle is 42.7097~amu, giving a frequency of oscillation between two CM particles of
\begin{align*}
\nu_{\text{bond}} = \frac{1}{2 \pi}  \sqrt{\frac{0.01587643}{\mu}}\text{~fs}^{-1} = 0.0043395872\text{~fs}^{-1}
\end{align*}
with reduced mass $\mu=21.35485$~amu and a numerical units conversion factor of 0.002577344. So, the CG bond oscillation period is $1/ \nu_{\text{bond}} \approx 230.44$~fs. Thus a time step of 5~fs resolves well the time evolution of the CG bond vibration forces, almost 50 time steps per oscillation. Meanwhile, the Lennard-Jones forces have a characteristic frequency of about
\begin{align*}
 \nu_{\text{LJ}} = \tau^{-1} = \sqrt{\frac{\epsilon}{m \sigma^2}}  \approx \frac{1}{2205}\text{~fs}^{-1}\;,
\end{align*}
showing that the LJ forces are extremely well-resolved for all choices of time step. Both $\nu_{\text{bond}}$ and $\nu_{\text{LJ}}$ suggest that our choice of $h=5$~fs is sufficiently small.

We used the molecular dynamics engine LAMMPS \cite{LAMMPS}, and implemented the integrators considered here using its Python interface (\texttt{fix python/move}) to the underlying data structures. The BAOAB, BBK and BBK$^*$ integrators are not available as packages in LAMMPS and were instead implemented using this Python interface. Although a G--JF option is available for the \texttt{langevin fix} command, the version of LAMMPS at the time of writing does not implement G--JF in same way as given in \cite{gjf}. Instead, it uses uniform random variables to approximate the Gaussian noise, in an effort to increase computational speed. However, such an approximation is only valid for small enough time steps \cite{gjf_approx}. As we are interested in the large time step regime, we implemented the original version of G--JF with Gaussian noise using the Python wrapper based on equations (20) and (21) in \cite{gjf}.

For each choice of $\gamma$ and $h$, 100 independent simulations were conducted to reduce the error associated to finite length simulations in approximating phase space averages. Each simulation was performed from an identical spatial configuration for approximately 250 ns. This starting configuration was obtained by an initialisation run using LAMMPS' \texttt{fix npt} command, which implements an MTK thermostat/barostat \cite{mtk}, for 100 ns with a temperature of 450 K and pressure set to 1 bar. All simulations used to benchmark the Langevin integrators were run in the NVT ensemble.

\section{Simulation Results}
The goal of our study is to understand how faithfully the different integrators reproduce relevant statistical averages of the coarse-grained model, particularly in the regime of large time steps and $\gamma$ values. To that end, numerical experiments were performed using a range of friction parameters and time steps. We examined the cases of $\gamma \in \{0.1, 0.01 ,0.001 ,0.0001\}$ and $h \in \{5,10,15,20,25,30,35\}$ with units of fs$^{-1}$ and fs, respectively.

In the simple case of a Brownian particle in a fluid, $\gamma$ represents the rate of collision of the Brownian particle with bath particles. So the dimensionless quantity $\gamma h$ gives a measure as to how many collisions occur over the length of the time step $h$.  Similarly here, the number $\gamma h$ determines the strength of the interaction between the system and the heat bath and is a fundamental quantity of the dynamics.

In applications of Langevin dynamics with atomistic models, friction rate parameters of the order of $.01$ fs$^{-1}$ or less are typically used, as they give a reasonable approximation of the experimental diffusion coefficients of small molecules. Indeed in \cite{bbk}, the authors used BBK in an atomistic water simulation as their benchmark test, and $\gamma$ was chosen quite small, $0.0000196 \text{ fs}^{-1}$. In \cite{lk_article}, the BAOAB method was tested and compared to other integrators based on its performance on an atomistic alanine dipeptide molecule in water with $\gamma = 0.001 \text{ fs}^{-1}$.

However, in CG simulation models, much of the magnitude of the inter-atomic forces shifts from the conservative to the stochastic terms, and a higher friction rate $\gamma$ may be needed to retain the same diffusivity. Alternatively, high friction rates are also used simply to improve numerical stability of MD simulations near particular conditions (for example, near phase transitions). Therefore, we consider here a relatively broad range of $\gamma$ values, $0.0001$ fs$^{-1}$ to $0.1$ fs$^{-1}$: given the choice of integration time step $h$ used in the following, the upper end of this interval may result in values of $\gamma h$ larger than $1$.

For smaller $\gamma$ values, i.e., $\gamma \le 0.001$ fs$^{-1}$, the three integrators become numerically unstable around $h\approx 38$ fs. Hence, 35 fs was chosen as the upper bound for our range of time step values. This stability limit for G--JF and BAOAB increases significantly for the largest choice of $\gamma=0.1 \text{ fs}^{-1}$ to slightly more than 50 fs. In contrast, this larger value for $\gamma$ seemed to not have as much of an effect on BBK$^*$ --- simulations still exhibited instability at 40 fs.

\subsection{Diffusive Behavior} \label{sec:diffusion}
\begin{figure}
\begin{subfigure}{0.49\textwidth}
  \centering
  \includegraphics[width=.9\textwidth]{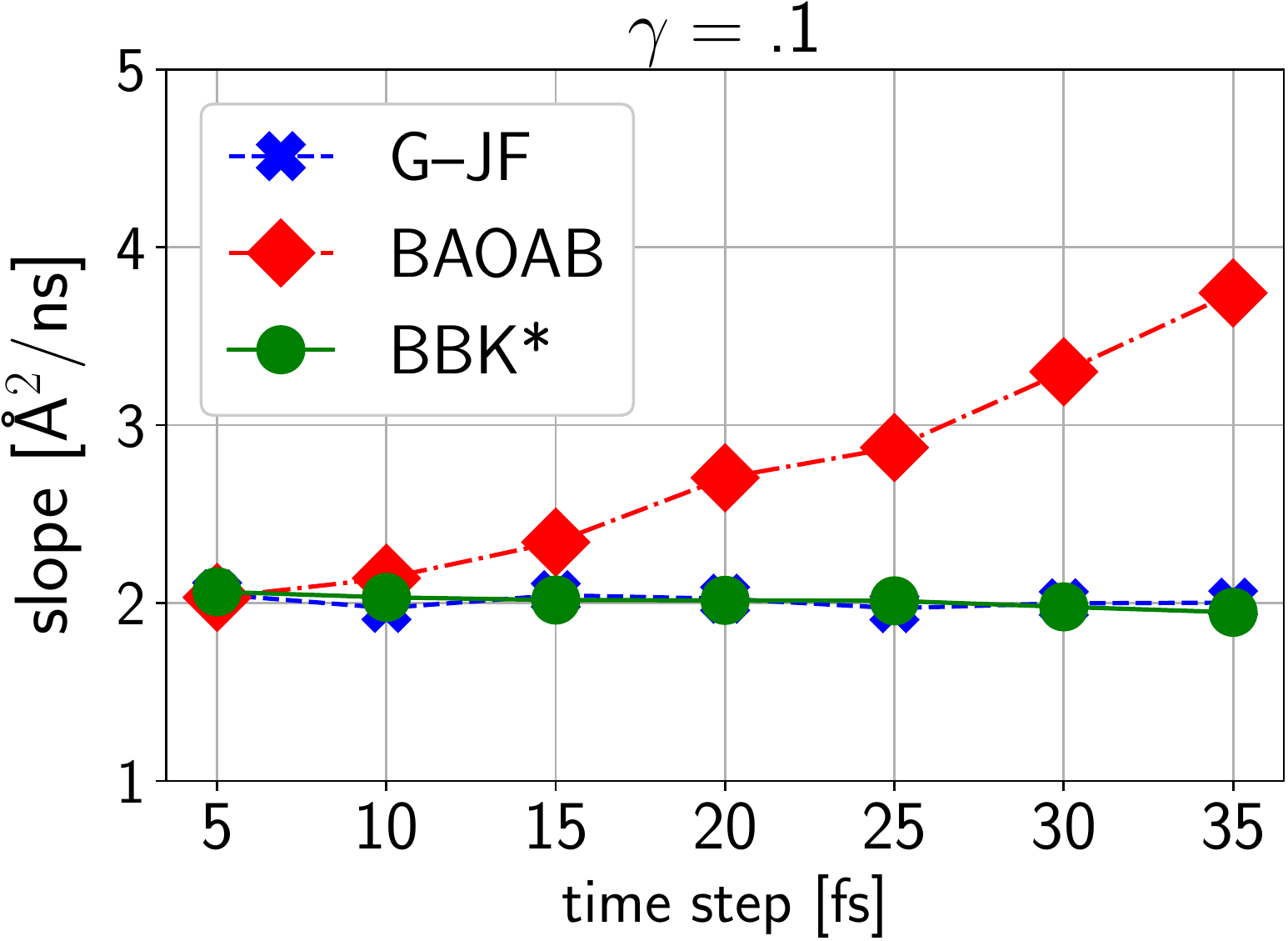} 
\end{subfigure}
\hfill
\begin{subfigure}{0.49\textwidth}
  \centering
  \includegraphics[width=.9\textwidth]{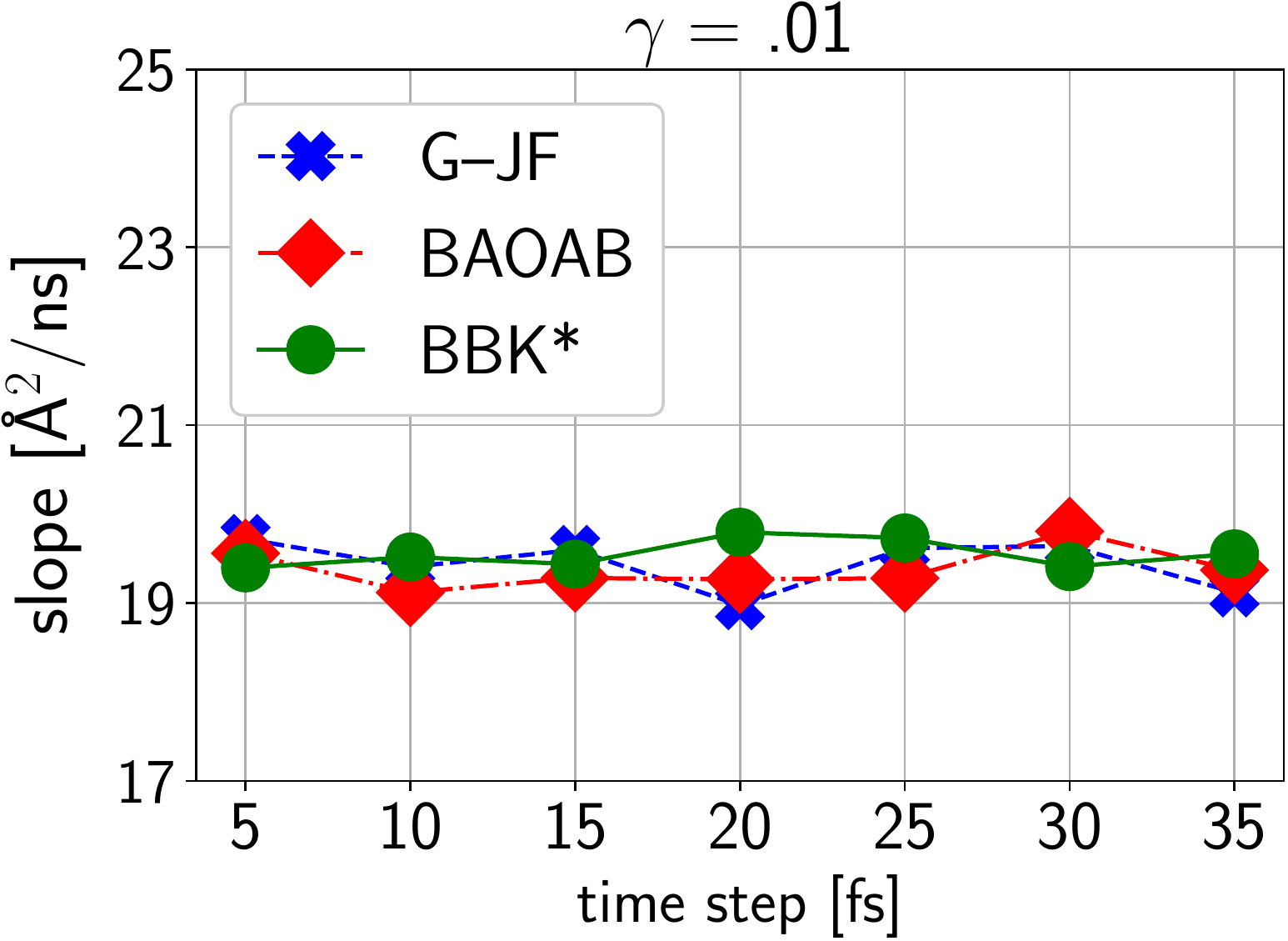} 
\end{subfigure}

\vspace{.5em}
\begin{subfigure}{0.49\textwidth}
  \centering
  \includegraphics[width=.9\textwidth]{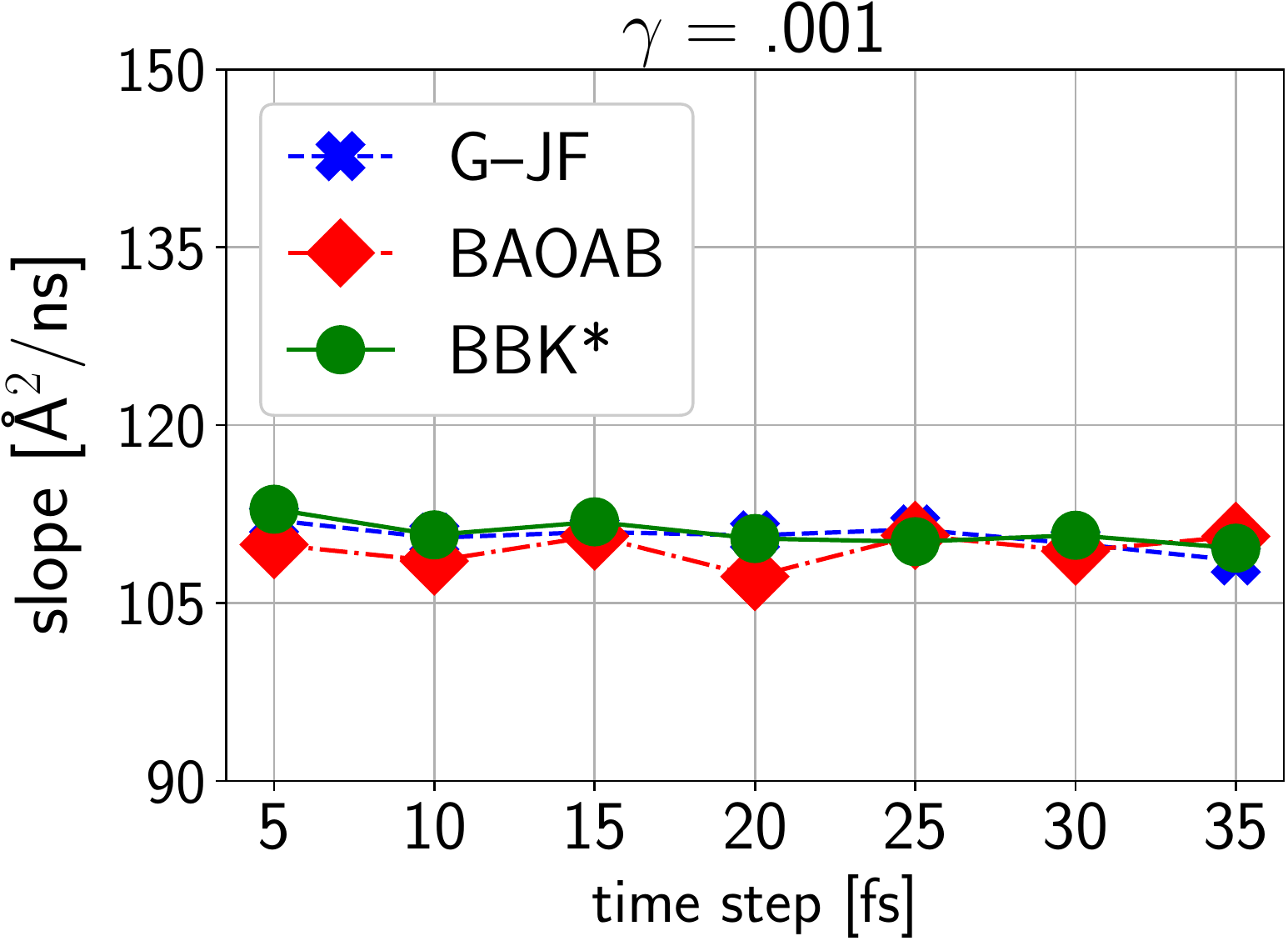} 
\end{subfigure}
\hfill
\begin{subfigure}{0.49\textwidth}
  \centering
  \includegraphics[width=.9\textwidth]{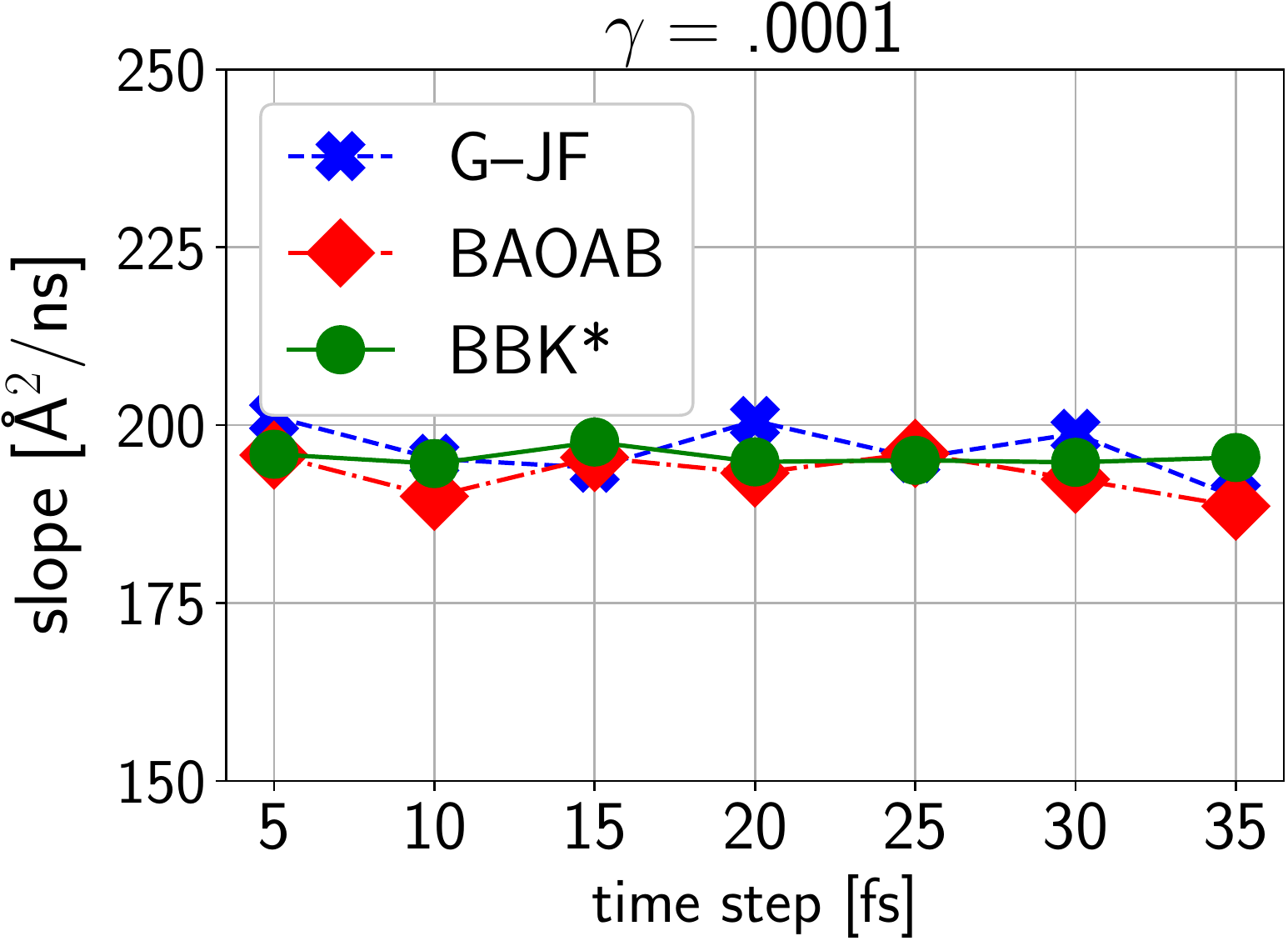} 
\end{subfigure}\\
\caption{Diffusion coefficients (with units of {\AA}$^2$/ns) were calculated from numerical simulation for different $\gamma$ values as a function of the time step $h$. The plot in the upper left displays the same kind of linear behavior for BAOAB as the method exhibits in the one-dimensional free particle. In the remaining cases, all three methods produce the same calculated slopes within statistical error as a function of $h$.} \label{fig:diffusion}
\end{figure}

To characterise diffusion, the mean squared displacement (MSD) of individual molecules was computed as a function of the simulation time. After each time step in the simulation, we computed the MSD of the center-of-mass for each polymer chain over that time step, using the LAMMPS command \texttt{compute msd}, and then averaged this result over all chains and added this to the same calculation from the previous step, i.e., we computed:
\[
\mathcal{D}_h(t_k):=\frac{1}{N} \sum_{i=1}^{N} |Q_i^{\text{CM}} (t_k) - Q_i^{\text{CM}}  (t_{k-1}) |^2 + \mathcal{D}_h(t_{k-1})\;,
\]
where $t_k=kh$, $1\le k \le N$, and $Q_i^{\text{CM}}$ represents the center of mass position for the $i$-th polymer chain ($i=1,..,128$). The center-of-mass drift of the entire system was subtracted from the MSD data at each time step before computing the MSD. The resulting MSD data was then block-averaged \cite{block_avg} over the 100 independent runs for each choice of $\gamma$ and $h$, yielding 5000 independent samples (each block was taken long enough to allow for correlations to die off). This was done for each of the three integrators. The MSD was observed to be linear with respect to time within statistical error. These linear plots were fitted with regression lines and the means of the slopes of these lines are plotted as a function of the time step $h$ in Figure~\ref{fig:diffusion}.
As discussed earlier, the diffusion coefficient for BAOAB in the case of the one-dimensional single particle with zero net external potential does not adhere to the Einstein diffusion relation (see Table~\ref{table:harmonic_numerical_stats}). In that same spirit, the diffusion coefficient increases linearly in the upper left plot of Fig.~\ref{fig:diffusion} as $h$ (and thus $\gamma h$) increases. The computed diffusion coefficients for the G--JF and BBK$^*$ integrators are relatively unchanged as a function of the time step, again, in line with the behavior in the simple one-dimensional case. This is a desirable property, as it provides evidence that using larger time steps with G--JF and BBK does not corrupt the system's diffusive behavior for any choice of $\gamma$. For $\gamma \le 0.01$ fs$^{-1}$, all integrators exhibit statistically similar diffusive behavior, giving confidence that the choice of integrator should not influence diffusion in this regime.

We would like to stress the fact that in this study, we observe only classical diffusion, unlike some previous studies. In \cite{polymer} and \cite{grest}, polymer melts with CG particles composed of three CH2 monomers, as considered here, were studied. Sub-diffusion was observed for the quantity $\mathcal{D}_h$ for simulation times up to and exceeding our simulation time of 250 ns (although the polymer lengths were at least double ours). Therefore we initially considered the possibility of anomalous diffusion during our simulation, however no such power law was observed. Further examination of error residuals with MSD data and regression lines did not provide evidence of non-linear relationships between MSD and simulation time.

\subsection{Configurational Averages}\label{sec:config_avgs}
An important quantity typically used in statistical thermodynamics is the radial distribution function (RDF). To discriminate the distributions of inter-molecular contacts from intra-molecular ones, we restricted the computation of the RDF to pairs of particles from distinct chains. The CM---CM distributions are here examined: because the system is mainly made up of CM particles, this RDF is the most fully sampled.

For each friction parameter $\gamma$ and time step $h$, the RDF for the intermolecular CM$-$CM particle pairing was calculated every 1000 time steps, and these calculated distributions were then averaged over time. This was done for each of the 100 independent simulations, followed by a final averaging over these 100 simulations in order to reduce sampling error.

\begin{figure}
\begin{subfigure}{0.49\textwidth}
  \centering
  \includegraphics[width=.9\linewidth]{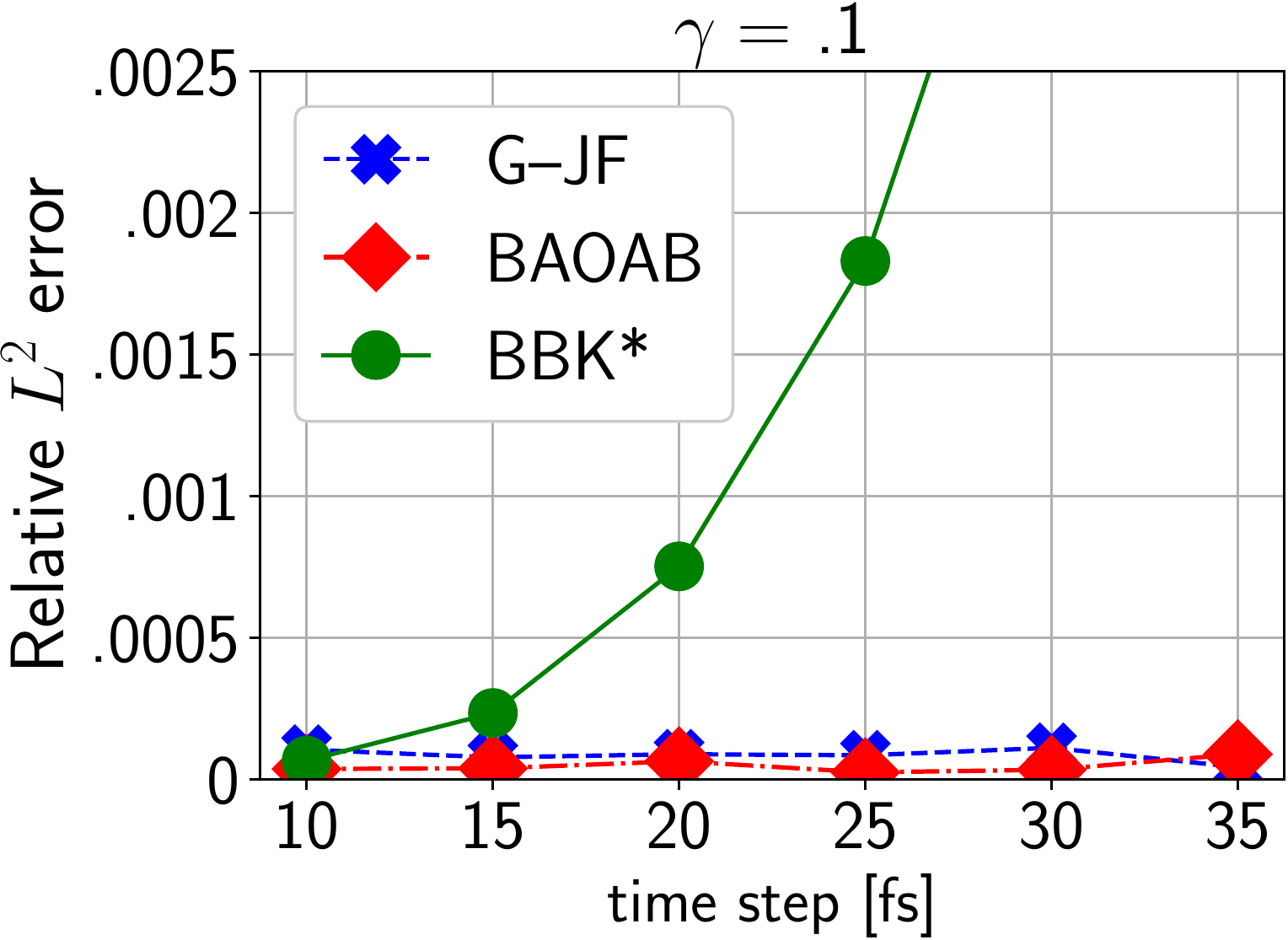} 
\end{subfigure}
\hfill
\begin{subfigure}{0.49\textwidth}
  \centering
  \includegraphics[width=.9\linewidth]{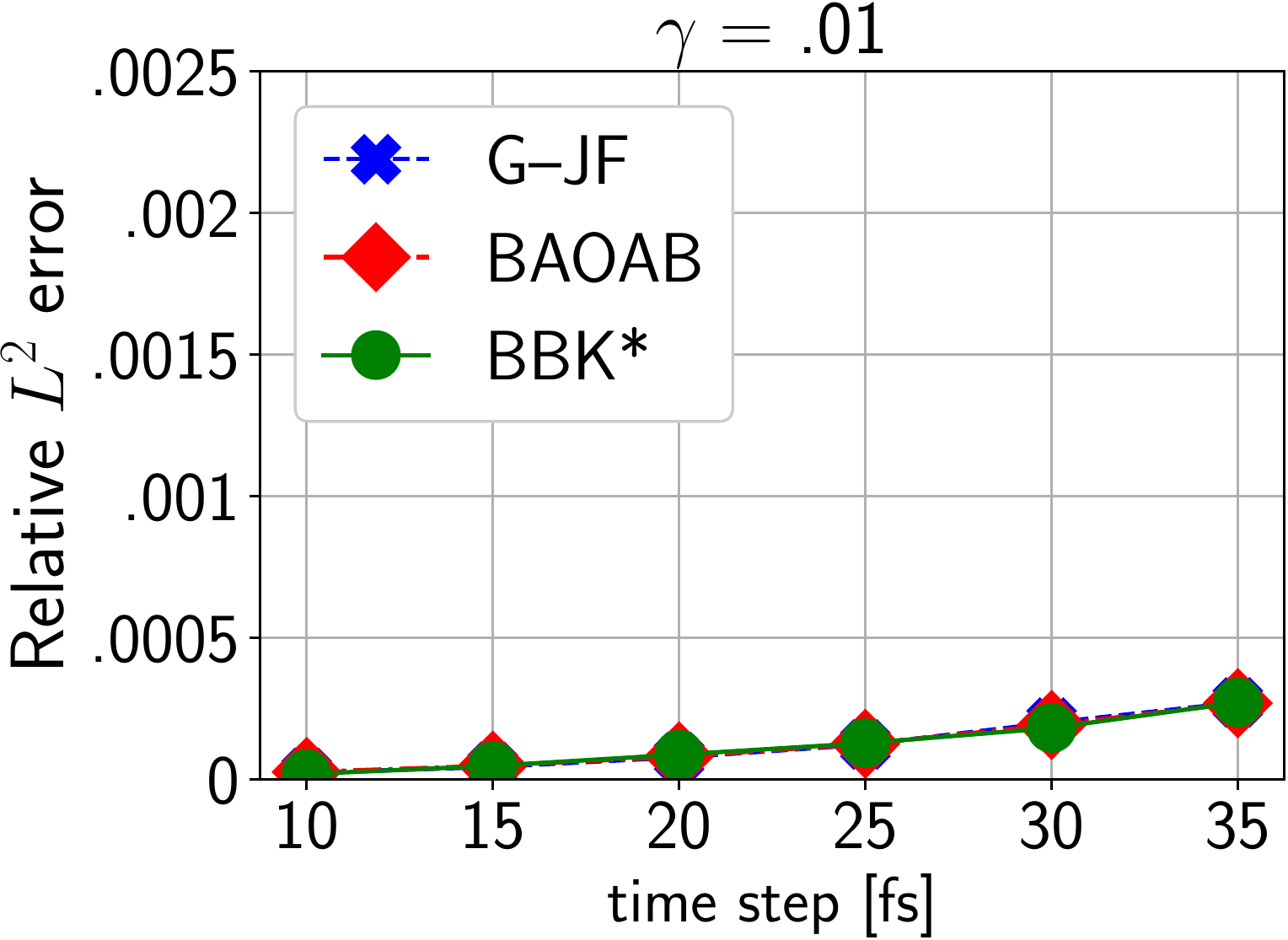} 
\end{subfigure}

\vspace{.5em}
\begin{subfigure}{0.49\textwidth}
  \centering
  \includegraphics[width=.9\linewidth]{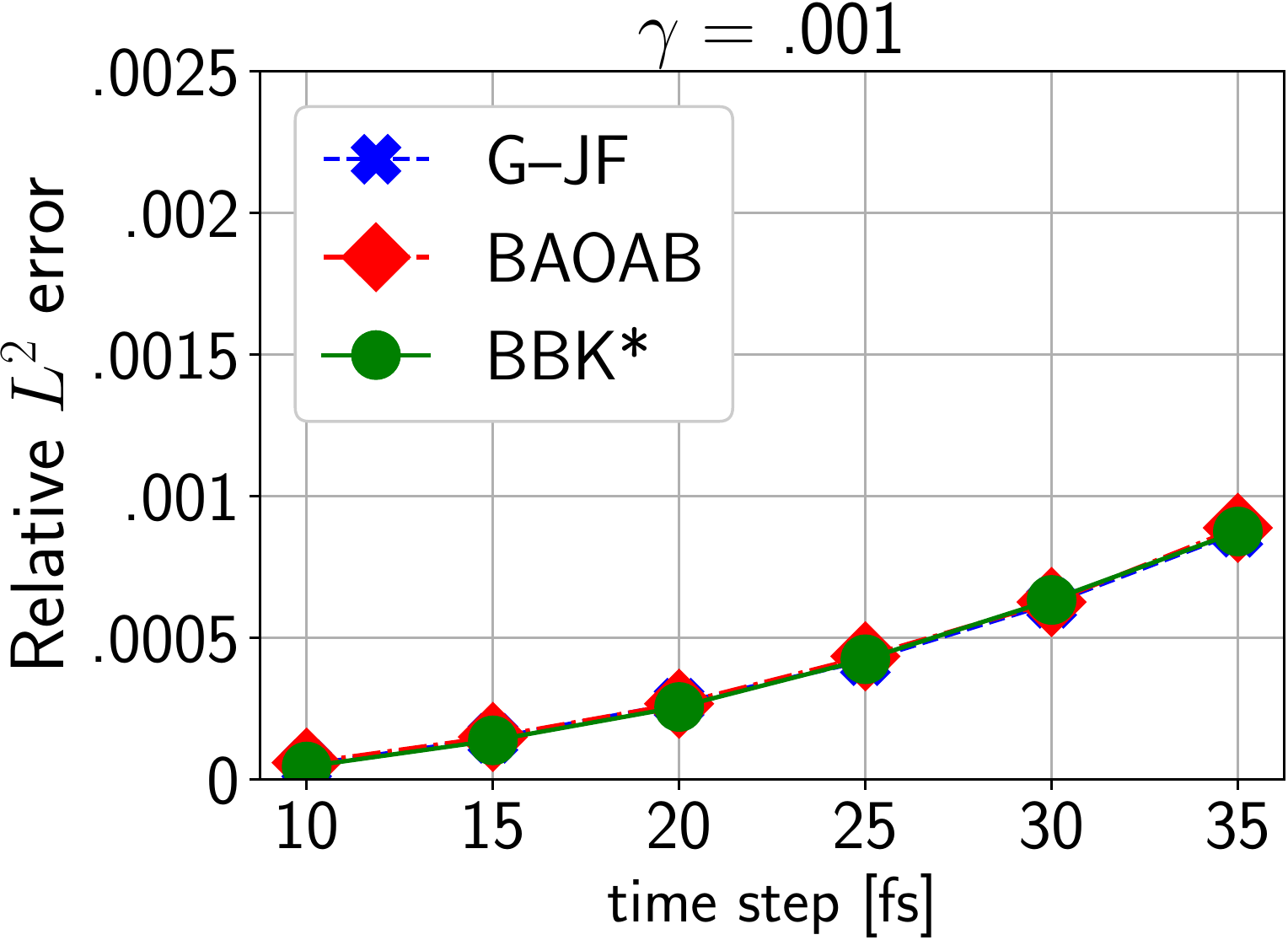} 
\end{subfigure}
\hfill
\begin{subfigure}{0.49\textwidth}
  \centering
  \includegraphics[width=.9\linewidth]{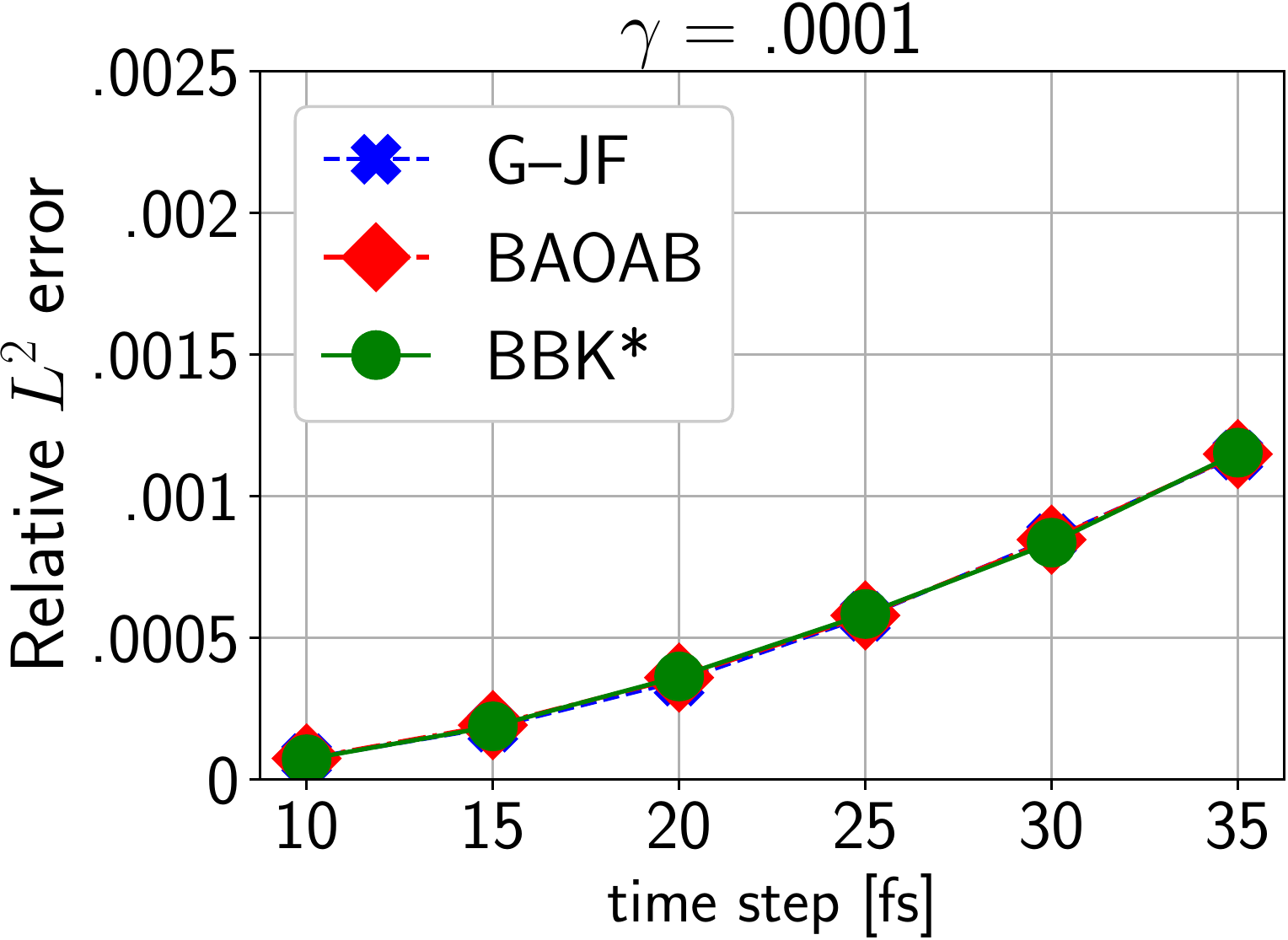} 
\end{subfigure}
\caption{Relative error of radial distribution functions for the three numerical schemes for four different values of $\gamma$. For the largest $\gamma$ value, we notice minimal change in error as the time step is increased for G--JF and BAOAB; but a more significant error for BBK$^*$. Although outside the range of the panel (a), for the largest integration step, we recorded an error of approximately 0.75\% in the RDF, which can make an impact on the quality of results. For smaller choices of $\gamma$, we observe behavior consistent with Hamiltonian dynamics: an increasing time step leads to increased error.} \label{fig:rdf}
\end{figure}

Lacking analytical expressions for the true RDF, computations conducted with time step $h=5$ fs are used as a reference solution. Given how small $h=5$ fs is relatively to the time scale of the overall processes, it is reasonable to assume that the true intermolecular CM$-$CM RDF for the CG system will be well-approximated by the one calculated for $h=5$ fs. For each $\gamma$ value, we calculate the reference RDF's: $g_{\text{G--JF}}^\gamma(r)$, $g_{\text{BAOAB}}^\gamma(r)$ and $g_{\text{BBK}^*}^\gamma(r)$. Then the $L^2$ relative differences between the reference RDFs and the RDFs obtained for the other choices of $h$ are computed and plotted as functions of $h$. These results are displayed in Figure~\ref{fig:rdf}.

\begin{figure}
  \centering
  \includegraphics[width=.5\linewidth]{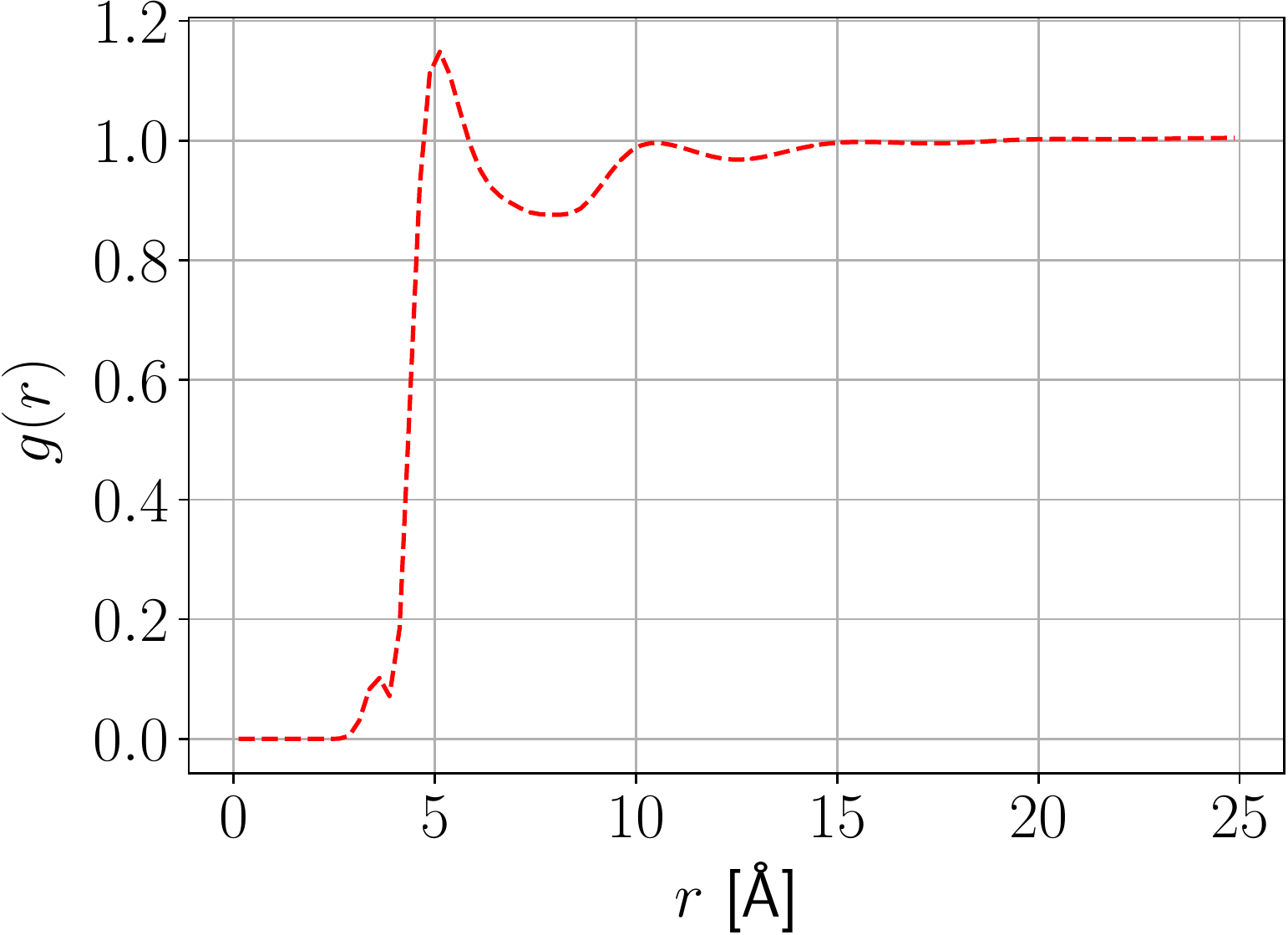} 
  \caption{For each $\gamma$ value and each integrator, the CM---CM radial distribution function for $h=5$ fs was calculated. These radial distribution functions were then used as a proxy for the true CG radial distribution function. For a given $\gamma$ value, we denote these reference RDF's as $g_{\text{G--JF}}^\gamma(r)$, $g_{\text{BAOAB}}^\gamma(r)$ and $g_{\text{BBK}^*}^\gamma(r)$. These were visually indistinguishable and so we only show $g_{\text{G--JF}}^{0.01}(r)$ here.
  }
\end{figure}
%

\begin{figure}
\centering
\includegraphics[width=.78\linewidth]{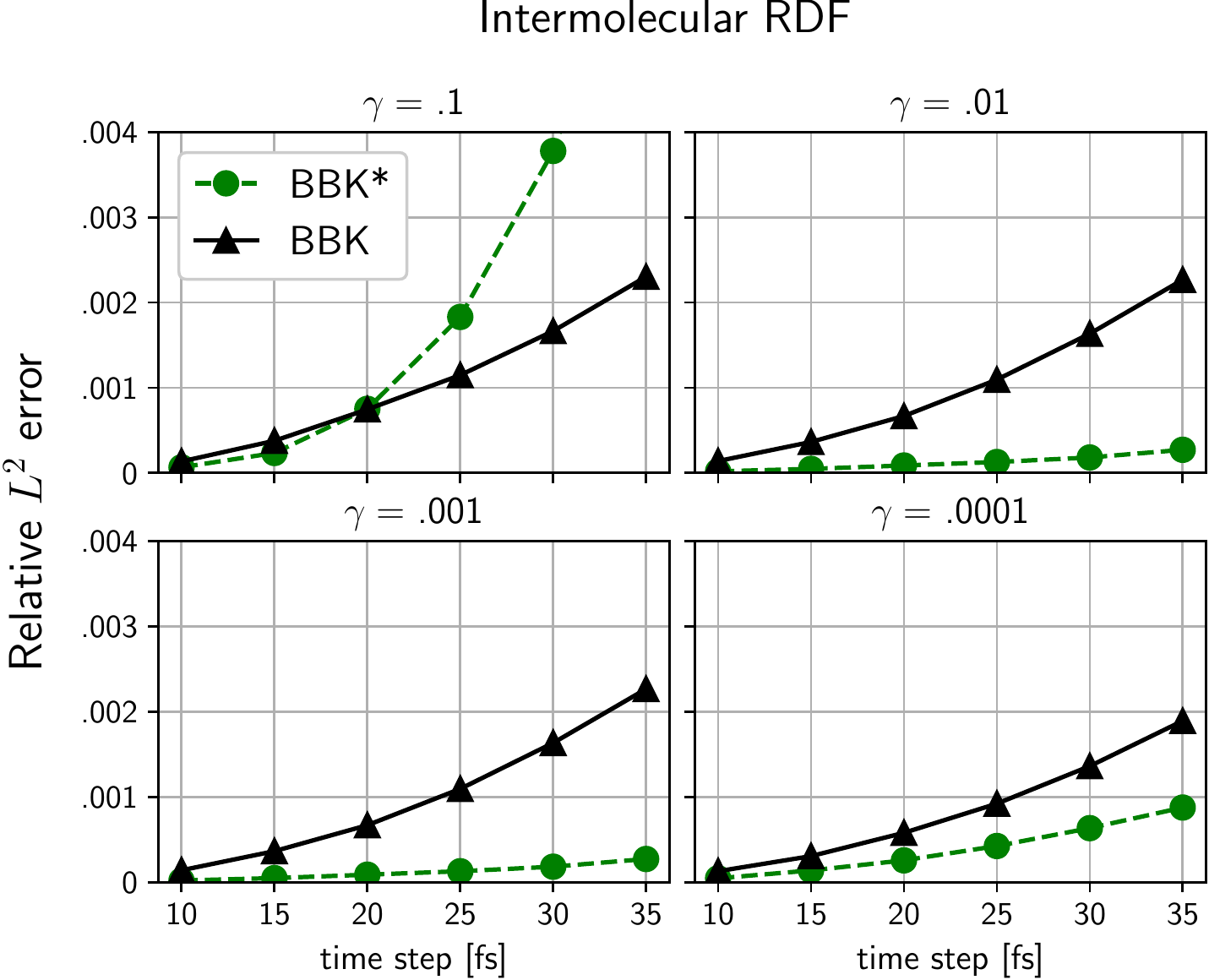} 

\vspace{2em}
\includegraphics[width=.78\linewidth]{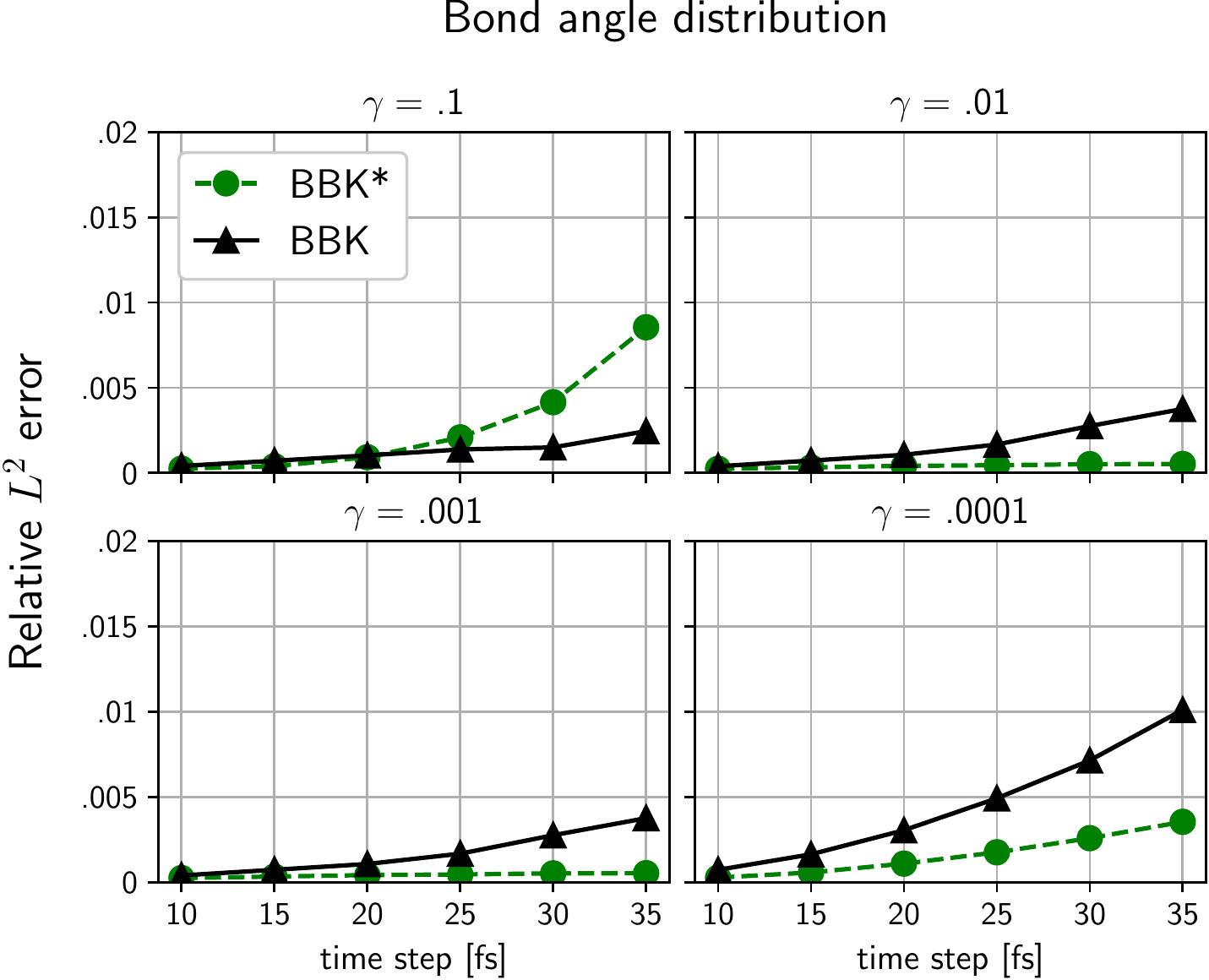} 
\caption{A comparison of radial and angle distribution errors for BBK and BBK$^*$.}\label{fig:bbk_vs_bbk*}
\end{figure}

In the large $\gamma$ regime, BBK$^*$ exhibits more and more deviation from its baseline RDF as $h$ is increased, while the other two methods remain unchanged. In fact, although not displayed in the plot, for $h=35$ fs, the BBK$^*$ error is larger than that of BAOAB and G--JF by an order of magnitude. If one considered here the original BBK formulation, the resulting error would be slightly lower than BBK* (Fig. \ref{fig:bbk_vs_bbk*}), but still much higher than BAOAB and G--JF. When $\gamma \le 0.01$ fs$^{-1}$, all three integrators (BAOAB, BBK$^*$ and G--JF) perform almost identically. This fact should be useful to the practitioner working in the small $\gamma$ regime when trying to strike a balance between diffusivity and adherence to the canonical distribution. Note also that for BBK*, this is a distinguishing property of the formulation used in this paper: the original BBK formulation has a larger error in this regime (Fig. \ref{fig:bbk_vs_bbk*}). Some intuition is available. If we consider the harmonic potential, the mean square position for BBK is given by $k_b T \omega^{-1}(1-\tfrac{h^2 \omega}{4m})^{-1}$; therefore in the simple linear case, the configurational statistics depend only on the size of the time step, much like what is seen in Fig \ref{fig:bbk_vs_bbk*}.

\begin{figure}
  \centering
  \includegraphics[width=.95\linewidth]{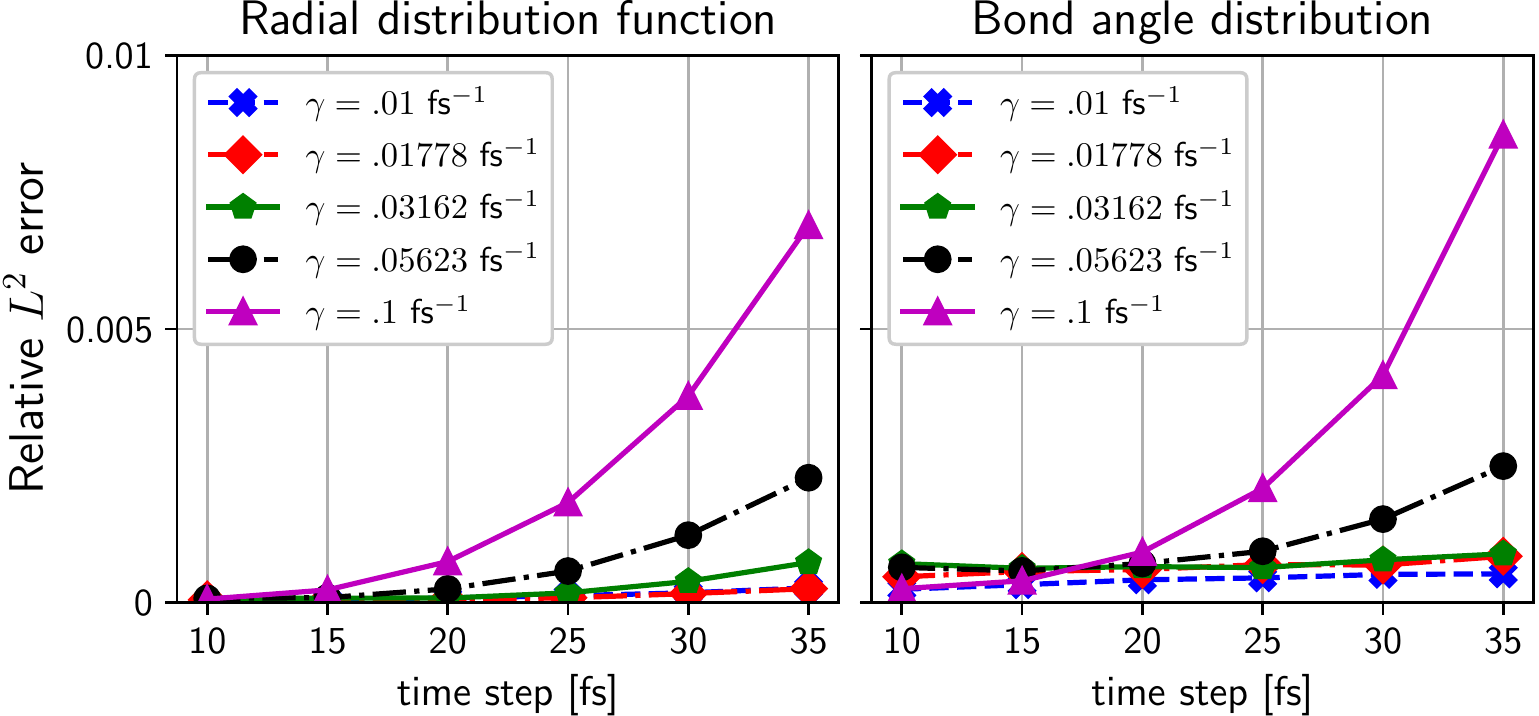} 
  \caption{Relative error of the radial and bond angle distribution for the BBK$^*$ method as $\gamma$ varies between $10^{-2}$ fs$^{-1}$ and $10^{-1}$ fs$^{-1}$, taking on the values $10^{-2}, 10^{-1.75},10^{-1.5},10^{-1.25}, 10^{-1}$ fs$^{-1}$. We see a smooth transition as $\gamma$ becomes smaller, not an abrupt phase transition, further suggesting dependence of the steady state distribution on the dimensionless quantity $\gamma h$.} \label{fig:bbk_transition}
\end{figure}

The divergence of the BBK$^*$ RDF from the other RDFs when $\gamma$ is largest, led us to question what happens when an intermediate $\gamma$ value is used. Figure~\ref{fig:bbk_transition} displays these results. A smooth transition of error occurs between $\gamma=.01$ fs$^{-1}$ and $\gamma=.1$ fs$^{-1}$. This suggests that the BBK$^*$ RDF exhibits a $\gamma h$ dependence that is not present in the other RDFs---which is not surprising, given that a similar behavior occurs in the simple 1-d harmonic oscillator case. Again, we stress that in most atomistic applications, $\gamma$ is simply not large enough for the $\gamma h$ dependence to be noticeable. However, this may no longer be the case for CG dynamics where the larger $\gamma$ regime becomes more relevant.

Another configurational quantity of interest for polymer chains is the bond angle distribution. Many properties of polymer melts, as well as their transition between liquid and solid phases, are determined by the propensity of the polymer chains to align. Failing to accurately reproduce the angle distribution can dramatically influence the accuracy of the simulation's thermodynamic properties.

\begin{figure}
\begin{minipage}{.49\textwidth}
  \centering
  \includegraphics[width=.9\linewidth]{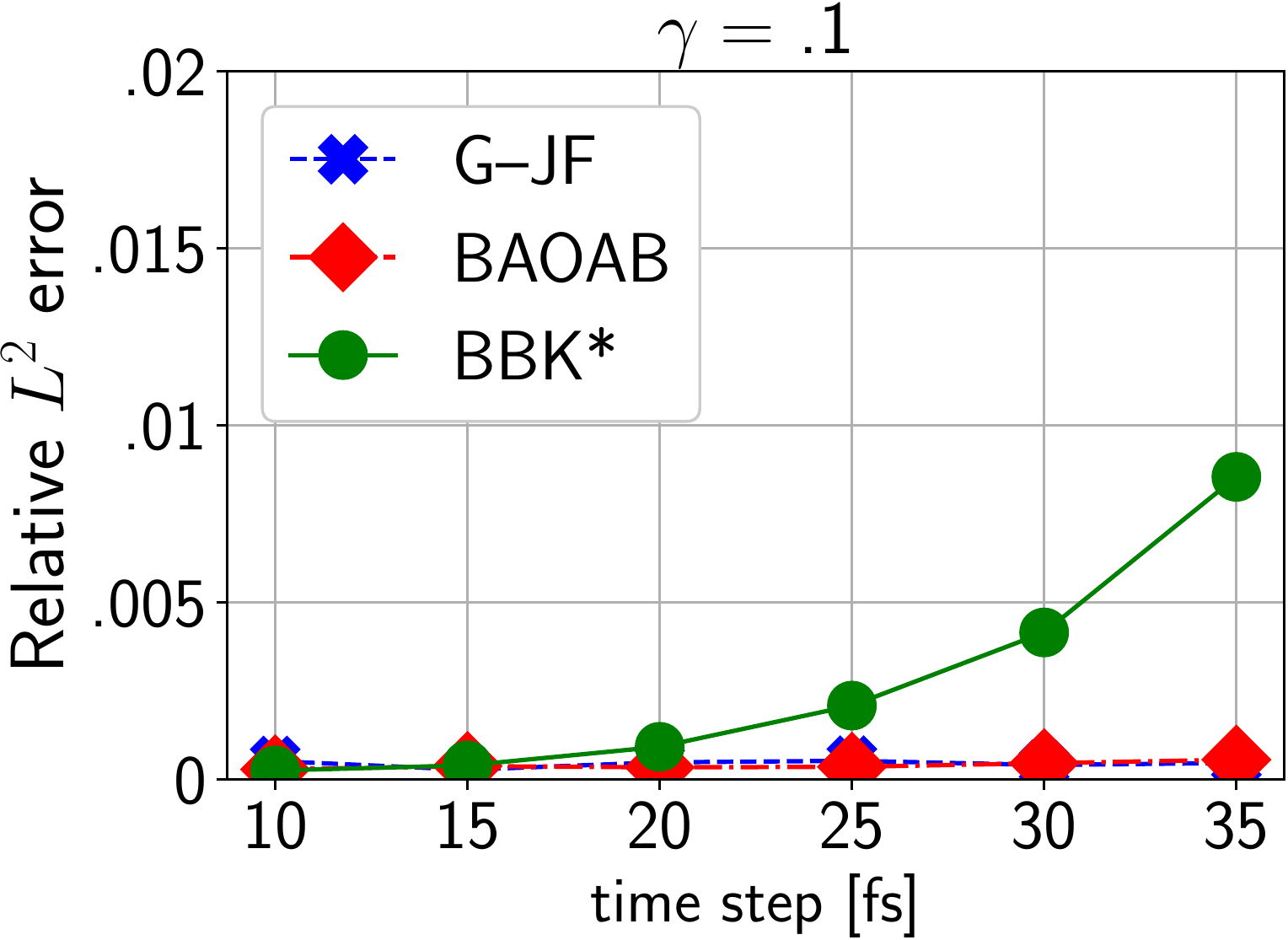} 
\end{minipage}
\hfill
\begin{minipage}{.49\textwidth}
  \centering
  \includegraphics[width=.9\linewidth]{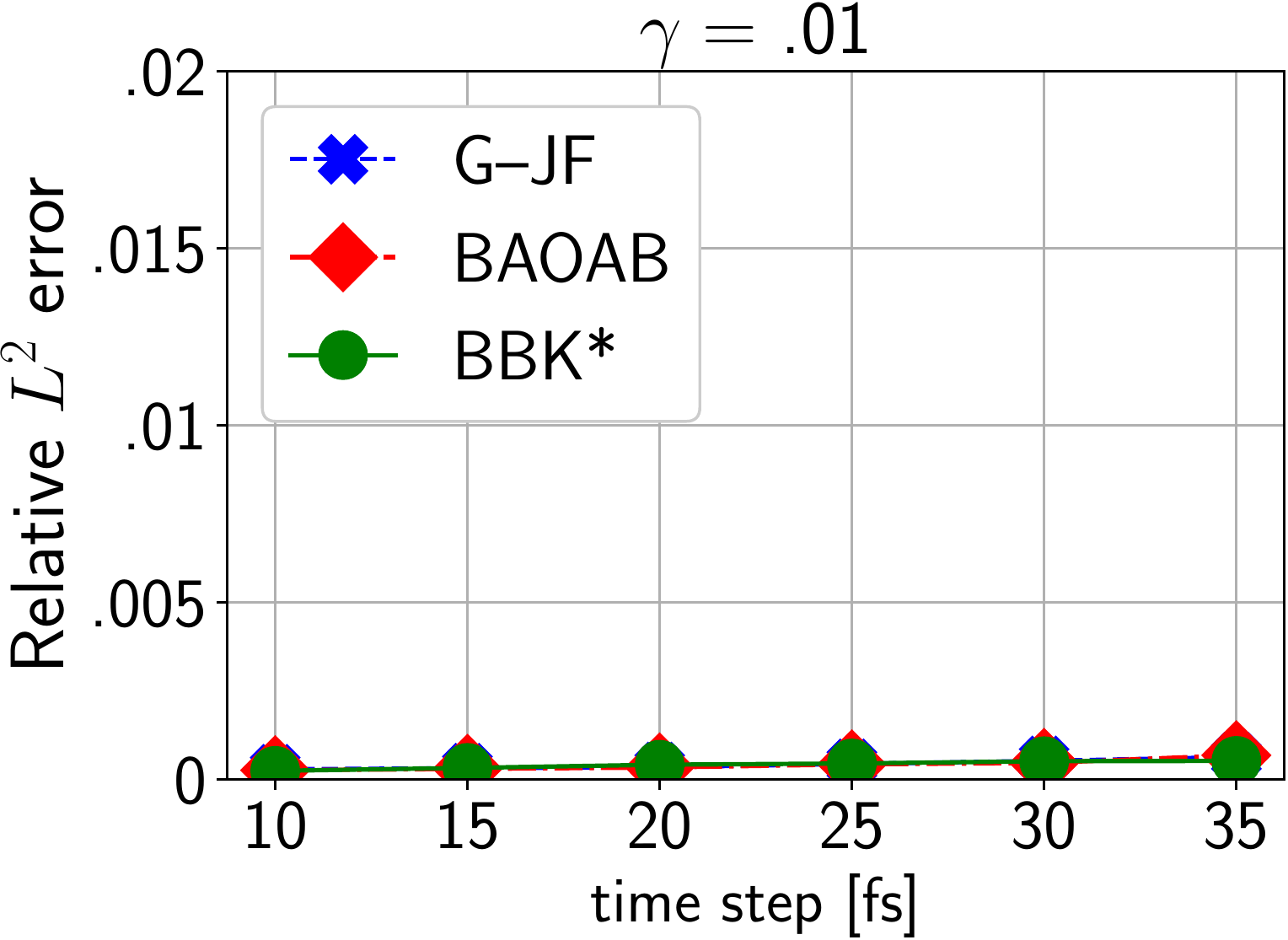} 
\end{minipage}

\vspace{.5em}
\begin{minipage}{.49\textwidth}
  \centering
  \includegraphics[width=.9\linewidth]{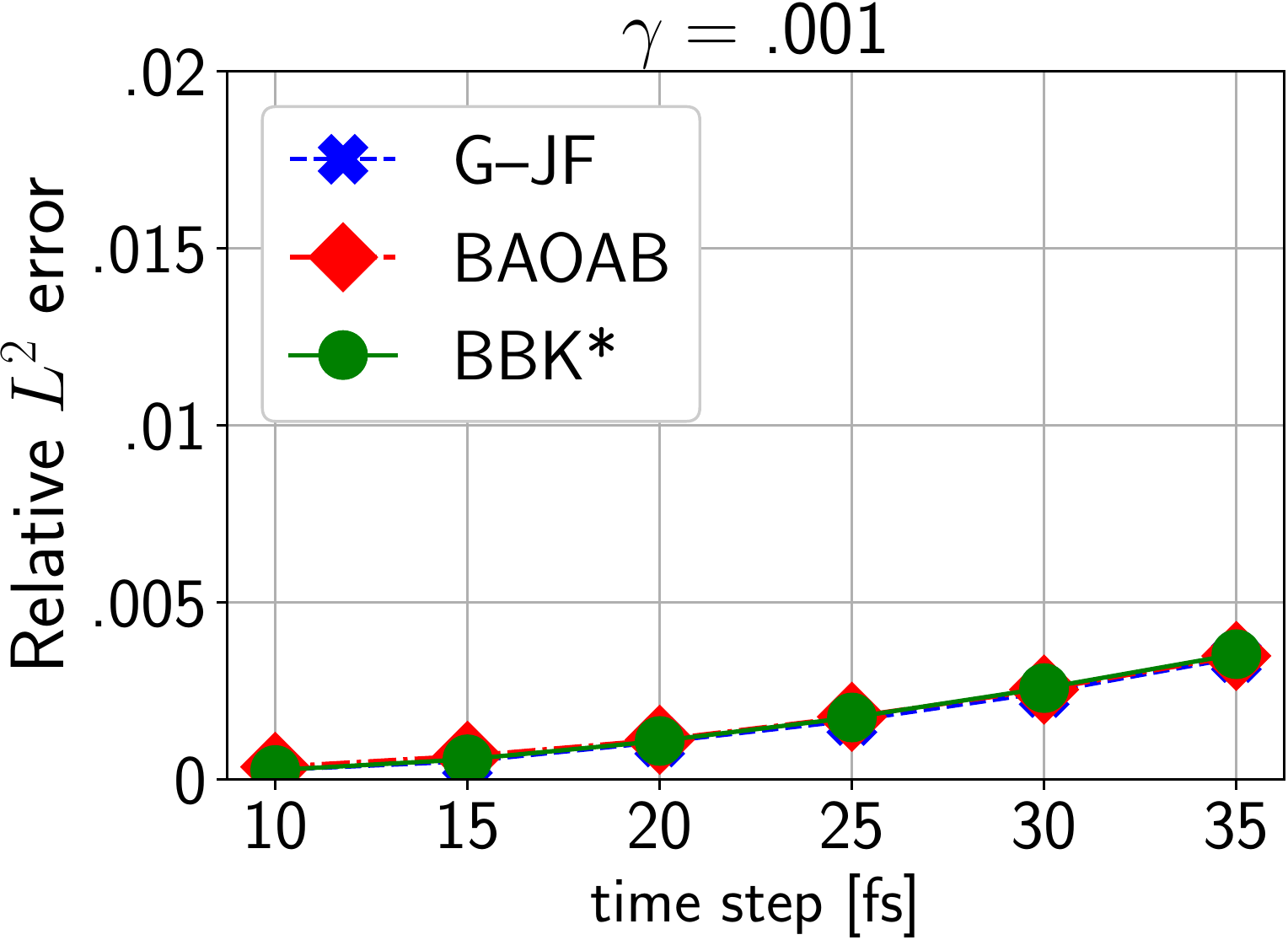} 
\end{minipage}
\hfill
\begin{minipage}{.49\textwidth}
  \centering
  \includegraphics[width=.9\linewidth]{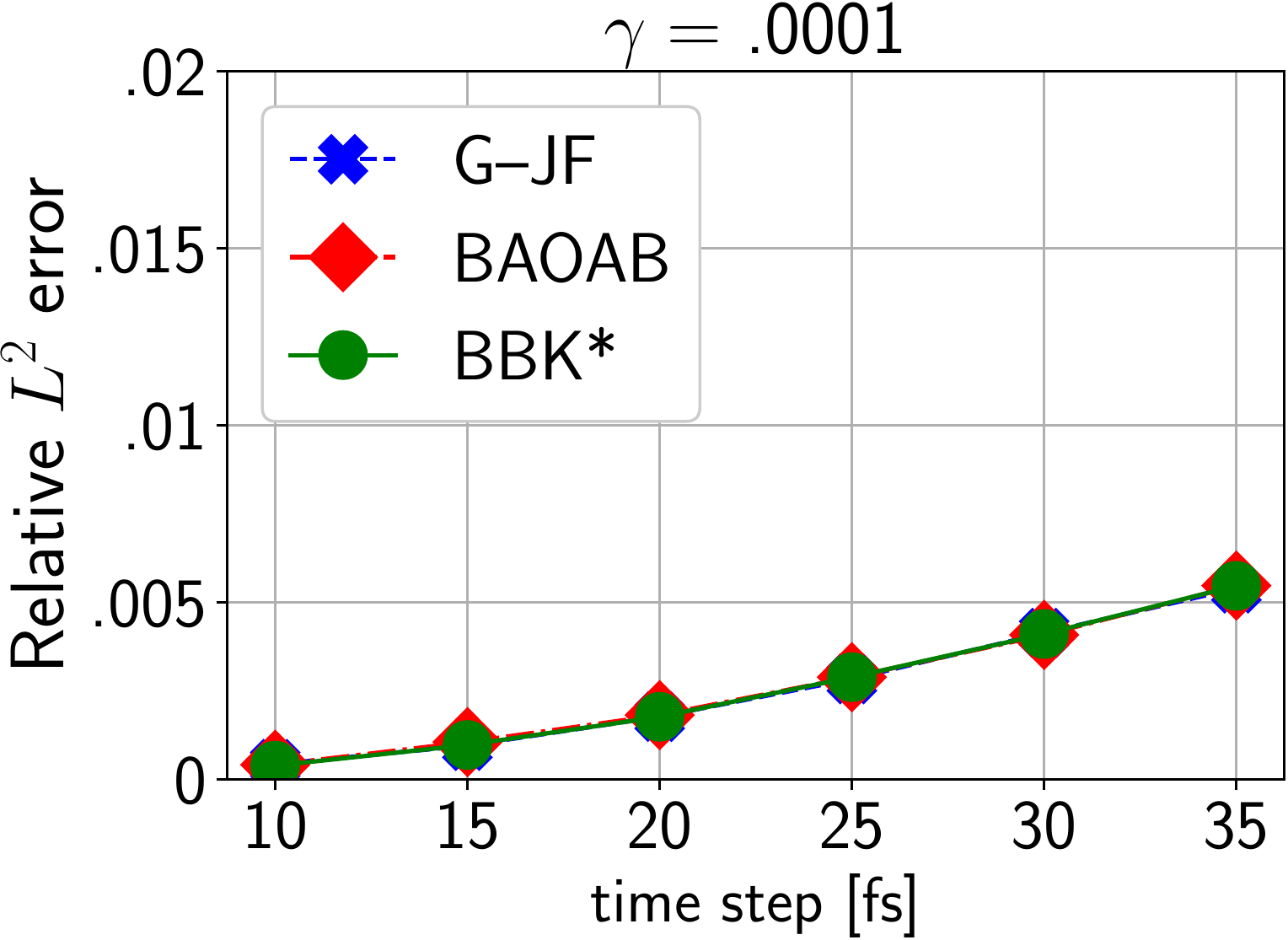} 
\end{minipage}
\caption{Relative error of the bond distribution for the three numerical schemes when $\gamma=0.1 \text{ fs}^{-1}, 0.01 \text{ fs}^{-1},0.001 \text{ fs}^{-1}, 0.0001 \text{ fs}^{-1}$. The reference is taken to be $h=5$ fs. As with the RDF, minimal change in error occurs when the time step is increased for G--JF and BAOAB, but for BBK$^*$ it is much larger. At 35 fs, BBK$^*$ experiences an almost 1\% error in its bond angle distribution. Convergence to Hamiltonian dynamics is seen in the bottom row.} \label{fig:bond_angle}
\end{figure}

Of the two possible triplets CM---CM---CT and CM---CM---CM, the latter was the best sampled and so was more amenable to statistical inference. For larger $\gamma$, G--JF and BAOAB produce minimal to no variation in error as $h$ is increased. Figure~\ref{fig:bond_angle} displays the results. BBK$^*$ again exhibits systematic errors for larger $\gamma$ values. The top row in Fig.~\ref{fig:bond_angle} indicates a large relative distortion of the bond angle distribution for BBK$^*$. This evidence may lead one to use caution in applying BBK$^*$ to liquid phase CG systems with larger $\gamma$ values, especially since this behavior for BBK$^*$ is also observed with the CM---CM intermolecular RDF. For smaller $\gamma$, that is as the system gets closer to pure Hamiltonian dynamics, again all three integrators perform similarly, indicating no significant preference of integrator for this regime. Table~\ref{table:harmonic_numerical_stats} showed that in a harmonic potential, the configurational statistics were dependent on $\gamma h$ for BBK$^*$. So it is not unreasonable to expect that a similar $\gamma h$ dependence for other configurational quantities may occur in more complicated situations when using BBK$^*$, as seen in our simulations.

We observe very good agreement of BBK$^*$ with G--JF and BAOAB for $\gamma \le .01$ fs$^{-1}$ in Figures \ref{fig:rdf} and \ref{fig:bond_angle}. However, for all choices of $\gamma$, the original BBK exhibits a clear trend: the relative error increases as the time step $h$ is increased.

\section{Conclusions}
In this paper we systematically studied three different Langevin integrators on a coarse-grained (CG) polymer melt. For the ideal cases of the Brownian motion and harmonic oscillator, key statistical properties were calculated analytically for each integrator, which provided guiding insights into diffusive and statistical behavior of realistic molecular systems. In particular, for pure Brownian motion, both BBK$^*$ and G--JF capture the true diffusive behavior exactly for all choices of $\gamma$ and $h$; but BAOAB is only approximate, with the diffusion coefficient depending on the dimensionless parameter $\gamma h$. This carried over to the CG-polymer simulation results. In Section \ref{sec:diffusion}, the calculated diffusion coefficient as a function of time step, $\mathcal{D}_h$, was found to be statistically independent of the time step $h$ for BBK$^*$ and G--JF for all $\gamma$ values, whereas BAOAB displayed the same type of linear behavior for $\gamma h$ values of $O(1)$ as in the free particle case.

The computational results indicate that G--JF is the only integrator among those considered here that describes equally well configurational distributions and diffusive behavior over all $\gamma$ choices. As expected, BBK$^*$ and BBK perform poorly for the largest choice of $\gamma$.
BAOAB samples equally well the configurational distributions, but exhibits a spurious dependence of the diffusivity on the integration time step near the high-friction regime. These conclusions have implications for CG MD simulations, where large ratios between friction and Hamiltonian forces are more frequently encountered than in atomistic ones. The evidence presented in this paper should be useful to the practitioner in supporting the use of the G--JF thermostat for CG simulations.

\section{Acknowledgments}
We would like to thank Richard Berger for technical assistance.
Part of this research was funded by the US Army Research Laboratory under contract number W911NF-16-2-0189. B. Seibold wishes to acknowledge support by NSF grant DMS--1719640. Calculations were carried out on Temple University's HPC resources and thus were supported in part by the National Science Foundation through major research instrumentation grant number 1625061.

\vspace{1.5em}
\bibliographystyle{plain}
\bibliography{bib}

\begin{thebibliography}{10}

\bibitem{lammps_source_code}
\uppercase{LAMMPS} source code, 2017.
\newblock \url{http://lammps.sandia.gov/}.

\bibitem{namd_source_code}
Sequencer.c, \uppercase{NAMD} source code, 2018.
\newblock \url{https://www.ks.uiuc.edu/Research/namd/}.

\bibitem{gjf_num_analysis}
S.~Anmarkrud, K.~Debrabant, and A.~Kv{\ae}rn{\o}.
\newblock General order conditions for stochastic partitioned
  \uppercase{R}unge--\uppercase{K}utta methods.
\newblock {\em BIT Numerical Mathematics}, 58(2):257--280, 2018.

\bibitem{gjf_CG_test}
E.~Arad, O.~Farago, and N.~Gr{\o}nbech-Jensen.
\newblock The \uppercase{G-JF} thermostat for accurate configurational sampling
  in soft-matter simulations.
\newblock {\em Isr. J. Chem.}, 56(8):629--635, 2016.

\bibitem{brannigan2006implicit}
G.~Brannigan, L.~Lin, and F.~Brown.
\newblock Implicit solvent simulation models for biomembranes.
\newblock {\em Eur. Biophys. J.}, 35(2):104--124, 2006.

\bibitem{bbk}
A.~Br{\"u}nger, C.~Brooks~III, and M.~Karplus.
\newblock Stochastic boundary conditions for molecular dynamics simulations of
  \uppercase{ST2} water.
\newblock {\em Chem. Phys. Lett.}, 105(5):495--500, 1984.

\bibitem{reverse_leapfrog}
K.~Burrage and G.~Lythe.
\newblock Accurate stationary densities with partitioned numerical methods for
  stochastic differential equations.
\newblock {\em SIAM J. Numer. Anal.}, 47(3):1601--1618, 2009.

\bibitem{cooke2005tunable}
I.~Cooke, K.~Kremer, and M.~Deserno.
\newblock Tunable generic model for fluid bilayer membranes.
\newblock {\em Phys. Rev. E}, 72(1):011506, 2005.

\bibitem{gjf_approx}
B.~D{\"u}nweg and W.~Paul.
\newblock Brownian dynamics simulations without \uppercase{G}aussian random
  numbers.
\newblock {\em Int. J. Mod. Phys. C}, 2(03):817--827, 1991.

\bibitem{block_avg}
H.~Flyvbjerg and H.~G. Petersen.
\newblock Error estimates on averages of correlated data.
\newblock {\em J. of Chem. Phys.}, 91(1):461--466, 1989.

\bibitem{goetz1998computer}
R.~Goetz and R.~Lipowsky.
\newblock Computer simulations of bilayer membranes: self-assembly and
  interfacial tension.
\newblock {\em J. Chem. Phys.}, 108(17):7397--7409, 1998.

\bibitem{gjf}
N.~Gr{\o}nbech-Jensen and O.~Farago.
\newblock A simple and effective \uppercase{V}erlet-type algorithm for
  simulating langevin dynamics.
\newblock {\em Mol. Phys.}, 111(8):983--991, 2013.

\bibitem{gjf_test}
N.~Gr{\o}nbech-Jensen, N.~Hayre, and O.~Farago.
\newblock Application of the \uppercase{G-JF} discrete-time thermostat for fast
  and accurate molecular simulations.
\newblock {\em Comput. Phys. Commun.}, 185(2):524--527, 2014.

\bibitem{hadley2010structurally}
K.~Hadley and C.~McCabe.
\newblock A structurally relevant coarse-grained model for cholesterol.
\newblock {\em Biophys. J.}, 99(9):2896--2905, 2010.

\bibitem{langevin_stabilization}
J.A. Izaguirre, D.P. Catarello, J.M. Wozniak, and R.~Skeel.
\newblock Langevin stabilization of molecular dynamics.
\newblock {\em J. Chem. Phys.}, 114(5):2090--2098, 2001.

\bibitem{izvekov2005multiscale}
S.~Izvekov and G.~Voth.
\newblock A multiscale coarse-graining method for biomolecular systems.
\newblock {\em J. Phys. Chem. B}, 109(7):2469--2473, 2005.

\bibitem{lk_article}
B.~Leimkuhler and C.~Matthews.
\newblock Robust and efficient configurational molecular sampling via
  \uppercase{L}angevin dynamics.
\newblock {\em J. Chem. Phys.}, 138(17):05B601\_1, 2013.

\bibitem{lk_article_2}
B.~Leimkuhler and C.~Matthews.
\newblock Robust and efficient configurational molecular sampling via
  \uppercase{L}angevin dynamics.
\newblock {\em J. Chem. Phys.}, 138(17):05B601\_1, 2013.

\bibitem{lk}
B.~Leimkuhler and C.~Matthews.
\newblock {\em Molecular Dynamics}.
\newblock Springer, 2015.

\bibitem{espresso}
Hans-J{\"o}rg Limbach, A.~Arnold, B.~Mann, and C.~Holm.
\newblock \uppercase{ESPR}es\uppercase{S}o --- an extensible simulation package
  for research on soft matter systems.
\newblock {\em Comput. Phys. Commun.}, 174(9):704--727, 2006.

\bibitem{martini}
S.~Marrink, H.~Risselada, S.~Yefimov, D.~Tieleman, and A.~De~Vries.
\newblock The \uppercase{MARTINI} force field: coarse grained model for
  biomolecular simulations.
\newblock {\em J. Phys. Chem. B}, 111(27):7812--7824, 2007.

\bibitem{NH_chains}
G.~J. Martyna, M.~Klein, and M.~Tuckerman.
\newblock Nos{\'e}--hoover chains: The canonical ensemble via continuous
  dynamics.
\newblock {\em J. Chem. Phys.}, 97(4):2635--2643, 1992.

\bibitem{mtk}
G.~L. Martyna, D.~Tobias, and M.~Klein.
\newblock Constant pressure molecular dynamics algorithms.
\newblock {\em J. Chem. Phys.}, 101(5):4177--4189, 1994.

\bibitem{murtola2009systematic}
T.~Murtola, M.~Karttunen, and I.~Vattulainen.
\newblock Systematic coarse graining from structure using internal states:
  Application to phospholipid/cholesterol bilayer.
\newblock {\em J. Chem. Phys.}, 131(5):08B601, 2009.

\bibitem{orsi2008quantitative}
M.~Orsi, D.~Haubertin, W.~Sanderson, and J.~Essex.
\newblock A quantitative coarse-grain model for lipid bilayers.
\newblock {\em J. Phys. Chem. B}, 112(3):802--815, 2008.

\bibitem{bbk88}
R.~Pastor, B.~Brooks, and A.~Szabo.
\newblock An analysis of the accuracy of \uppercase{L}angevin and molecular
  dynamics algorithms.
\newblock {\em Mol. Phys.}, 65(6):1409--1419, 1988.

\bibitem{polymer}
B.~Peters, K.~M. Salerno, A.~Agrawal, D.~Perahia, and G.~Grest.
\newblock Coarse-grained modeling of polyethylene melts: effect on dynamics.
\newblock {\em J. Chem. Theory Comput.}, 13(6):2890--2896, 2017.

\bibitem{dpd_chapter}
I.~Pivkin, B.~Caswell, and G.~Karniadakis.
\newblock {\em Reviews in Computational Chemistry, Chapter 2}, volume~27.
\newblock John Wiley \& Sons, 2011.

\bibitem{LAMMPS}
S.~Plimpton.
\newblock Fast parallel algorithms for short-range molecular dynamics.
\newblock {\em J. Comput. Phys.}, 117(1):1--19, 1995.

\bibitem{SHAKE}
Jean-Paul Ryckaert, G.~Ciccotti, and H.~Berendsen.
\newblock Numerical integration of the cartesian equations of motion of a
  system with constraints: molecular dynamics of n-alkanes.
\newblock {\em J. Comput. Phys.}, 23(3):327--341, 1977.

\bibitem{grest}
K.~M. Salerno, A.~Agrawal, D.~Perahia, and G.~S. Grest.
\newblock Resolving dynamic properties of polymers through coarse-grained
  computational studies.
\newblock {\em Phys. Rev. Lett.}, 116(5):058302, 2016.

\bibitem{SS}
T.~Schneider and E.~Stoll.
\newblock Molecular-dynamics study of a three-dimensional one-component model
  for distortive phase transitions.
\newblock {\em Phys. Rev. B}, 17(3):1302, 1978.

\bibitem{baoab_test}
X.~Shang, M.~Kr{\"o}ger, and B.~Leimkuhler.
\newblock Assessing numerical methods for molecular and particle simulation.
\newblock {\em Soft Matter}, 13(45):8565--8578, 2017.

\bibitem{shelley}
J.~Shelley, M.~Shelley, R.~Reeder, S.~Bandyopadhyay, and M.~Klein.
\newblock A coarse grain model for phospholipid simulations.
\newblock {\em J. Phys. Chem. B}, 105(19):4464--4470, 2001.

\bibitem{sdk}
W.~Shinoda, R.~DeVane, and M.~Klein.
\newblock Multi-property fitting and parameterization of a coarse grained model
  for aqueous surfactants.
\newblock {\em Mol. Simulat.}, 33(1-2):27--36, 2007.

\bibitem{tuckerman}
M.~Tuckerman.
\newblock {\em Statistical Mechanics and Molecular Simulations}.
\newblock Oxford University Press, 2008.

\bibitem{tuckerman_respa}
M.~Tuckerman, B.~Berne, and G.~Martyna.
\newblock Reversible multiple time scale molecular dynamics.
\newblock {\em J. Chem. Phys.}, 97(3):1990--2001, 1992.

\bibitem{wang_skeel}
W.~Wang and R.~Skeel.
\newblock Analysis of a few numerical integration methods for the
  \uppercase{L}angevin equation.
\newblock {\em Mol. Phys.}, 101(14):2149--2156, 2003.

\end{thebibliography}

\vspace{1.5em}
\end{document}